\documentclass{aastex63}

\def\ang{\AA}

\def\gapprox{\lower.4ex\hbox{$\;\buildrel >\over{\scriptstyle\sim}\;$}}
\def\lapprox{\lower.4ex\hbox{$\;\buildrel <\over{\scriptstyle\sim}\;$}}
\def\ref#1{\par\noindent\hangindent1cm {#1}}

\begin{document}

\title{Self-Organized Criticality in Stellar Flares}

\correspondingauthor{Markus J. Aschwanden}
\email{aschwanden@lmsal.com}

\author{Markus J. Aschwanden}

\affiliation{Solar and Stellar Astrophysics Laboratory (LMSAL),
 Palo Alto, CA 94304, USA}

\and

\author{Manuel G\"udel}

\affiliation{University of Vienna, Astrophysics Department,
 T\"urkenschanzstrasse 17, 1180 Vienna, Austria}

\begin{abstract}
Power law size distributions are the hallmarks of nonlinear
energy dissipation processes governed by self-organized
criticality. Here we analyze 75 data sets of stellar flare
size distributions, mostly obtained from the {\sl
Extreme Ultra-Violet Explorer (EUVE)} and the {\sl Kepler}
mission. We aim to answer the following questions
for size distributions of stellar flares: 
(i) What are the values and uncertainties of power law
slopes? (ii) Do power law slopes vary with time ? (iii) 
Do power law slopes depend on the stellar spectral type? 
(iv) Are they compatible with solar flares? (v) Are they
consistent with self-organized criticality (SOC) models? 
We find that the observed size distributions of stellar flare 
fluences (or energies) exhibit power law slopes of 
$\alpha_E=2.09\pm0.24$ for optical data sets observed with Kepler. 
The observed power law slopes do not show 
much time variability and do not depend on the stellar 
spectral type (M, K, G, F, A, Giants). 
In solar flares we find that background subtraction
lowers the uncorrected value of $\alpha_E=2.20\pm0.22$ 
to $\alpha_E=1.57\pm0.19$. 
Furthermore, most of the stellar flares are temporally
not resolved in low-cadence (30 min) Kepler data, which
causes an additional bias. Taking these two biases 
into account, the stellar flare data sets are consistent 
with the theoretical prediction $N(x) \propto x^{-\alpha_x}$ 
of self-organized criticality models, i.e., $\alpha_E=1.5$. 
Thus, accurate power law fits require automated 
detection of the inertial range and background subtraction,
which can be modeled with the generalized Pareto distribution, 
finite-system size effects, and extreme event outliers. 
\end{abstract}


\section{Introduction}

Statistical
observations of stellar flares started in the years of 1966-1972 with
ground-based spectroscopy in optical wavelengths (Moffett 1974;
Moffett and Bopp 1976; Lacy et al.~1976), 
and was then extended to {\sl extreme ultra-violet (XUV)} wavelengths
with space-based instruments such as EXOSAT, the {\sl Extreme
Ultra-Violet Explorer (EUVE)}, Chandra, ROSAT, ASCA, the Hubble 
Space Telescope (HST), and the Newton {\sl X-ray Multi-Mirror Mission 
(XMM-Newton)} 
(Collura et al.~1988; Pallavicini et al.~1990; Osten and Brown 1999;
Robinson et al.~1999; Audard et al.~1999, 2000; Kashyap et al.~2002).
A handicap of space-based observations is the sparse time coverage
of stellar flares, which severely restricts the measurement of power 
law slopes of flare energy distributions to small scale-free
(i.e., inertial) energy ranges.  Consequently, 
sophisticated modeling attempts have been developed to extract
power law slopes from small flare samples with $\approx 5-15$ 
events only (Audard et al.~1999, 2000; Kashyap et al.~2002; 
G\"udel et al.~2003; Arzner and G\"udel 2004; Arzner et al.~2007; 
Stelzer et al.~2007).
One motivation of power law modeling of stellar flare size
distributions is the fact that a power law slope of 
$\alpha \approx 2$ represents a critical value that decides whether 
the total energy of a size distribution diverges at the largest 
flares (if $\alpha < 2$), 
or whether it diverges at the smallest flares (so-called nanoflares) 
(if $\alpha > 2$). The critical divergence to infinity, however, 
goes away if one restricts the energy integral to observed energy 
ranges only, rather than extrapolating to un-observed size ranges.
The extrapolation of the power law to unobserved energies that are 
many orders of magnitude smaller remains questionable
(Benz and Krucker 2002). Nevertheless, extrapolation of nanoflare
size distributions from small to large flares provides almost sufficient
energy flux to heat the corona (Audard et al.~1999, 2000;
Kashyap et al.~2002; G\"udel et al.~2003; Arzner and G\"udel 2004;
Stelzer et al.~2007).

This all changed with the advent of the Kepler mission
(Borucki et al.~2010), providing a stellar flare catalog
with 162,262 events today, which due to the large statistics allows now
very accurate determinations of the power law slope of flare
size distributions (Maehara et al.~2012; Shibayama et al.~2013;
Aschwanden 2015; Wu et al.~2015; Davenport 2016; 
{\bf Van Doorsselaere et al.~2017}; Yang and Liu 2019). 
Indeed, fitting a
power law distribution to the entire Kepler data set yields a
very small statistical error, i.e., $\alpha=1.817\pm0.005$,  
which appears to be much smaller than systematic errors
due to truncation effects, instrumental sensitivity limitations, 
incomplete sampling, background subtraction effects,
arbitrary inertial ranges, and finite-system size effects
(Aschwanden 2015). Theoretically and observationally, 
there is evidence for two types of
white-light emission during flares (non-thermal beam heating
versus thermal continuum emission), which implies a prompt
as well as a delayed component, with possibly different
power law slopes of their size distributions.

On the modeling side we have to ask whether the value
of the power law slope depends on the stellar spectral
type, the age and size of the star, the rotation rate,
and the time variability (quiescent versus flaring time 
intervals). In this study we investigate some of
these effects, in order to improve the observed value
of the power law slope and to test theoretical 
{\sl Self-organized criticality (SOC)} models.
One modification of the standard SOC model is the
Dragon-King hypothesis (Sornette 2009;
Sornette and Ouillon 2012), which
suggests that the most extreme events in a statistical
distribution may belong to a different population, and
thus may be generated by a different physical mechanism,
in contrast to the strict power law behavior of standard
SOC models. Occasional evidence was found for such
a Dragon-King concept in solar and in stellar
flare data sets with large statistics (Aschwanden 2019, 2021).

The content of this paper includes a compilation of 
observations of stellar flares (Section 2, Tables 1-2),
examples of size distribution modeling (Section 3), 
statistical results (Section 4), discussion of relevant 
literature (Section 5), and conclusions (Section 6). 

\section{Observations of Stellar Flares}

We compile 20 studies that contain statistics of stellar flare 
radiated energies $E$, in form of occurrence frequency 
distributions $N(E) \propto
E^{-\alpha}$, with fitted power law slopes $\alpha$ (see 
Tables 1 and 2). Stellar flares are generally defined in terms 
of temporary increases of the observed flux above the pre-flare and 
post-flare quiescent level. The flare energy $E$ is often
estimated from the fluence or total counts (in a given wavelength), 
which is defined as the time integral of the flux (intensity
or luminosity) 
over the flare duration. Some data sets contain flares from a 
single star only, while other data sets combine flare events 
from different stars of the same stellar spectral type. 
In cases where the cumulative size
distribution is given (with slope $\beta$), we convert it to
the slope $\alpha=(\beta+1)$ of the differential size distribution.
For the uncertainty $\sigma_\alpha$ of the power law slope we use
the generic relationship $\sigma_\alpha=\alpha/\sqrt{n_{ev}}$
(Clauset 2009), where $n_{ev}$ represents the
number of flare events that contribute to the linear regression
fit of the power law slope. 
In Tables 1 and 2 and in this Section we summarize the instruments, 
the spectral types of the stellar classification, the number of flare events
per data set, the physical parameter used in the energy definition
(count rate, total counts, time-integrated radiated energy, or
peak flux rate), and the power law slopes of energies with 
uncertainties. 

Observations of 409 stellar flares have been obtained 
during 469 hours from 8 UV Ceti-type (cool dwarf) flare stars
(CN Leo, UV Cet, Wolf 424 AB, YZ CMi, EQ Peg, EV Lac, AD Leo,
YY Gem) in optical wavelengths (U, B, V band,
$\lambda=3370-5550$ \ang ), observed at the
McDonald Observatory (Moffett 1974; Lacy et al.~1976).
The flare energies $E$ were calculated from the observed
(time-integrated) fluences (or luminosities) for each star 
separately, yielding cumulative power law distributions 
with slopes in the range of $\beta=0.43-1.00$, which we convert to
differential size distributions with $\alpha=\beta+1=1.43-2.00$
{\bf (Moffett 1974; Lacy et al.~1976).}
A similar data set of 21 flares occurring on 13 different dMe 
stars was observed in soft X-rays with EXOSAT
(in soft X-ray energies of $0.05-2$ keV), yielding a
power law distribution of the soft X-ray peak flux 
$\alpha_P=1.52$ (Collura et al.~1988), and 
$\alpha_P=1.7$ (Pallavicini et al.~1990). These two studies
are the only ones that used a peak flux (in units of energy
per time), rather than a total (time-integrated) energy as
quoted in all other observations in this study. 
An analysis of 16 RS CVn binary star systems (G, F, K-types)
was undertaken with {\sl Extreme Ultraviolet Explorer (EUVE)} 
photometry, yielding a power law slope of energies $\alpha_E
=1.60$, where the flare energy is proportional to the 
time-integrated counts (Osten and Brown 1999). 
Photometric observations of the dM4.5e flare star YZ CMi using
the {\sl Hubble Space Telescope} in the ultraviolet wavelength 
range of ($\lambda=1600-3200$ \ang ) detected 54 flare events
and yielded an energy distribution with a power law-like slope of
$\alpha=2.25$ (Robinson 1999). Further observations with EUVE
targeted two active solar analogs, 47 Cas and EK Dra, gathering
28 flares and a power law slope of $\alpha \approx 2.2\pm0.2$
for the flare energies estimated from the total counts 
(Audard et al.~1999). Twelve data sets of cool dwarf stars
i.e., HD 2726, 47 Cas, EK Dra, $\kappa$ Cet (1994, 1995),
AB Dor, $\epsilon$ Eri, GJ 411, AD Leo, CN Leo (1994, 1995), 
were observed with EUVE, leading to power law slopes in
the range of $\alpha \approx 1.53-2.39$ (Audard et al.~2000).
One study focused on relatively weak flares from cool dwarfs
FK Aqr (dM2e type), V1054 Oph (M3Ve type), and AD Leo (M3 V type), 
and found power law slopes of $\alpha \approx$ 2.60, 2.74, and 2.03-2.32, 
using EUVE deep survey data (Kashyap et al.~2002).
Observations with EUVE concentrate on the cool dwarf AD Leo
in several studies (Kashyap et al.~2002; G\"udel et al.~2003;
Arzner and G\"udel 2004), applying different statistical models,
which converge to similar values for the power law slope in the
range of $\alpha \approx$ 2.25-2.30. Two other studies detected
soft X-ray emission from stellar flares using the XMM-Newton 
telescope, performing observations in the Taurus molecular cloud 
(Arzner et al.~2007; Stelzer et al.~2007). 

On March 7, 2009, the Kepler space telescope was launched
(Borucki et al.~2010), 
which was designed to survey the stars and planets in the 
Milky Way galaxy, in particular the exoplanets that orbit
around other stars than our Sun. It turned out that the
study of the light curves reveals numerous stellar flare
events, superimposed on the relatively slow rotational
modulation. At the time of writing, the Kepler flare
catalog contains 3420 flare stars and 162,262 flare events
(Yang and Liu 2019). {\bf The number of detected flares
strongly depends on the detection algorithm. For instance,
the study of Van Doorsselaere et al.~(2017)
detected 16,850 flares on 6662 stars out of a total of
188,837 in the Kepler field-of-view during Q15.} 
We provide a compilation of Kepler
flare statistics in Table 2, based on the studies of
Maehara et al.~2012; Shibayama et al.~2013; Balona 2015;
Aschwanden 2015, 2019, 2021;  
Wu et al.~2015, Davenport 2016, and Yang and Liu 2019). 
These new measurements
from Kepler represent an enormous breakthrough in the
statistical accuracy of stellar flares, which
enables us to determine the power law
distributions of stellar flare energies in different 
stellar types. However, one particular property is the wavelength
range of Kepler observations, which is in the optical wavelength
regime ($\lambda=4300-8900$ \ang ), roughly corresponding
to white-light flares observed on the Sun. We have to keep this
optical wavelength bias in mind when we compare with observations
in soft X-rays and EUV (EXOSAT, EUVE, XMM-Newton). 
Also, there may be a possible wavelength dependence
in the related power law slopes.
The flare events
have been detected with Kepler using different methods that
are described in Davenport (2016) and Yang and Liu (2019)  
and are subject to various biases and false-positive
signals. Examples of such erroneous signals are periodic
flux modulations from RR Lyrae, pulsating giants, eclipsing
binaries, or fluctuations from $\delta$ Scuti and $\gamma$
Doradus stars. Size distributions of flare-radiated
energies are gathered for flares from the same star
(given by the KID identification number in Table 2), 
as well as for different stellar spectral
types (A, F, G, K, M-type, Giants) separately.
  
\section{	Size Distribution Modeling 		}

The primary goal of our analysis is a deeper physical
understanding of size (or occurrence frequency)
distributions of stellar flares, which we can study now
thanks to the revolutionary {\sl Kepler} mission data
(Yang and Liu 2019),
with larger statistics and superior data quality than 
previously achieved with solar flare data. 
Solar data make up for a small subset of stellar data only. 
We expect that size distributions of solar and
solar-like stars should be similar 
(besides their amplitude that varies with 
the magnetic cycle), while it is
an open question whether other stellar types have
different size distributions than solar-like stars.
In the following we describe state-of-the-art power law
fitting methods of size distributions, applied here to 
data from the
Kepler stellar flare catalog. For more methodical details 
see also Aschwanden (2011, Section 7).

\subsection{	Differential Size Distribution 	 	}

A {\sl size distribution} of events, $N(x) dx$, also called 
a {\sl differential occurrence frequency distribution}, 
can ideally be approximated by a power law 
as a function of some size parameter $x$, quantified  
by four parameters ($x_1, x_2, n_0, \alpha_x$),
\begin{equation}
        N(x) dx = n_0 \ x^{-\alpha_x} dx \ ,
        \qquad x_1 \le x \le x_2 \ ,
\end{equation}
where $x_1$ and $x_2$ are the lower and upper bounds of the 
power law inertial (or scale-free) range, $\alpha_x$ is the power law
slope, and $n_0$ is a normalization constant. Uncertainties
in the calculation of the power law slope mostly result from
the arbitrary choice of the fitting range $[x_1, x_2]$, which
can be largely eliminated by a generalized power law function
that includes under-sampling at the lower end and finite 
system size effects at the upper end of the inertial range 
$[x_1, x_2]$.

\subsection{	Thresholded Power Law Distribution 	 	}

The ideal power law function (Eq.~1) can be generalized
with an additional ``shift'' parameter $x_0$ (with respect to $x$),
also called {\sl Lomax distribution} (Lomax 1954),
{\sl Generalized Pareto distribution} (Hosking and Wallis
1987), or {\sl Thresholded power law size distribution}
(Aschwanden 2015),
\begin{equation}
        N(x) dx = n_0 \left( x_0 + x \right)^{-\alpha_x} dx \ ,
        \qquad x_1 \le x \le x_2 \ ,
\end{equation}
with the normalization constant $n_0$.
This additional parameter $x_0$ accomodates three different
features: truncation effects due to incomplete sampling of
events below some threshold (if $x_0 > 0$), incomplete
sampling due to instrumental sensitivity limits (if $x_0 > 0$),
or subtraction of event-unrelated background (if $x_0 < 0$),
as it is common in astrophysical data sets (for details
see Aschwanden 2015). The differential size distribution 
(Eq.~2) fitted to the flare energies of the Kepler events 
is shown in Fig.~(1b) for the Pareto model, in Fig.~2b for
the finite-system model, and in Fig.~3b for the extreme
event model.

\subsection{	Cumulative Size Distribution		}

While Eqs.~(1) and (2) represent a {\sl differential size
distribution} $N_{diff}(x)$, it is statistically more
advantageous to employ {\sl cumulative size distributions} $N_{cum}(>x)$,
especially for small data sets and near the upper cutoff,
where we deal with a small number of events per bin.
We will use cumulative size distributions that include all events
accumulated above some size $x$, such as for the 
{\sl thresholded power law size distribution},
\begin{equation}
        N_{cum}(>x) = \int_{x}^{x_2} n_0
        ( x + x_0 )^{-\alpha_x} dx
        = 1 + (n_{ev}-1) \left( {(x_2+x_0)^{1-\alpha_x}-
	(x  +x_0)^{1-\alpha_x} \over (x_2+x_0)^{1-\alpha_x}-
	(x_1+x_0)^{1-\alpha_x} } \right) \ ,
\end{equation}
with $n_{ev}$ the number of events.  
We show an example of a size distribution that contains the counts
of events per bin $n_{cts}(x)$ in Fig.~(1a), the corresponding
differential size distribution $N_{diff}(x)=n_{cts}(x)/\Delta x$ in Fig.~(1b),
and the corresponding cumulative size distribution
$N_{cum}(>x)$ in Fig.~(1c).
The event count histogram (Fig.~1a) defines the inertial range $[x_0, x_2]$,
the minimum $(x_1)$, and maximum value $(x_2)$ of the size parameter $x$,
with a peak in the event count histogram at $x_0$ (Fig.~1a), which provides
a suitable threshold definition, because incomplete
sampling of small values $x < x_0$ is manifested by the drop of
detected events on the left side of the peak $x_0$. This definition of
a threshold $x_0$ has been proven to be very useful for characterizing
data sets with incomplete or sensitivity-limited sampling (Aschwanden 2015).
This provides also a definition for the inertial range $[x_0, x_2]$
in the fitting procedure.
 
\subsection{	Rank-Order Plot				}

An equivalent method to calculate the cumulative size distribution is the
{\sl rank-order plot}.
If the statistical sample is rather small, in the sense that no reasonable
binning of a histogram can be done, either because we do not have
multiple events per bin or because the number of bins is too small to
represent a distribution function, we can create a rank-order plot.
A rank-order plot is essentially an optimum adjustment to small
statistics, by associating a single bin to every event. From an event
list of a parameter $x_i, i=1, ..., n_x$,
(e.g., the energies of $n_x$ flares), 
which is generally not sorted,
we have first to generate a rank-ordered {\bf or sorted 1-D array}
by ordering the events according to increasing size,
\begin{equation}
        x_1 \le x_2 \le ... \le x_j \le ... \le x_n \ , \quad j=1, ..., n \ .
\end{equation}
The bins are generally not equidistant, neither on a
linear nor a logarithmic scale, defined by the difference between subsequent
values of the ordered series $x_j$,
\begin{equation}
        \Delta x^{bin}_j = x^{bin}_{j+1} - x^{bin}_j \ .
\end{equation}
In a rank-ordered sequence of $n_x$ events, the probability for the largest
value is $1/n_x$, for events that are larger than the second-largest event it is
$2/n_x$, and so forth, while events larger than the smallest event occur
in this event list with a probability of unity. Thus, the cumulative
frequency distribution is simply the reversed rank order,
\begin{equation}
        N_{cum}(>x_j) = ( n_x + 1-j ) \ , \qquad j=1,...,n_x \ ,
\end{equation}
and the distribution varies from $N_{cum}(>x_1)=n_x$ for $j=1$, to
$N_{cum}(>x_n)=1$ for $j=n_x$.

We can thus plot a cumulative frequency distribution with $N_{cum}(>x_j)$
on the y-axis versus the size $x_j$ on the x-axis. 
The cumulative distribution functions $N_{cum}(>x)$ (Eq.~6) of our 3 models
for the example of Kepler flares
is displayed in form of histograms in
Figs.~1c, 2c, and 3c, while the 5 highest
rank-ordered bins $N_{cum}(>x_j)$ are marked with diamonds 
in Figs.~1c, 2c, and 3c.

The distribution is
normalized to the number of events $n_x$,
\begin{equation}
        \int_{x_1}^{x_n} N(x) dx = N_{cum}(>x_1)=n_x \ .
\end{equation}
We overlay rank-order bins in the cumulative size distribution
of Fig.~(1c) (shown with black diamond symbols), for the 5
most extreme events of this size distribution. While the best fit
of the {\sl generalized Pareto distribution} yields a good fit in
the inertial range between $x_0=6.41 \times 10^{33}$ and 
$x \approx 10^{37}$ erg, we notice a significant deviation 
in the range above $x \gapprox 10^{37}$ erg, 
which we will model in the following. The self-consistency 
between the differential and the cumulative size distribution
is very high, with almost identical power law slopes of
$\alpha_{diff}=1.863 \pm 0.005$ (Fig.~1b) and 
$\alpha_{cum}=\beta_{cum}+1=1.877 \pm 0.005$ (Fig.~1c),
differing by $\lapprox 0.7\%$ only. {\bf This statistical 
uncertainty (or formal error) appears to be much smaller 
than systematic errors due to truncation effects, instrumental 
sensitivity limitations, incomplete sampling, background 
subtraction effects, arbitrary inertial ranges, and finite-system 
size effects (Aschwanden 2015).}

\subsection{    Finite-System Size Effects              }

The size distribution can be uniquely approximated with a classical
power law function, if the lower bound ($x_1$) and upper bound
($x_2$) are well-defined. In practice, however, the lower bound
is flattened by undersampling, detection thresholds, or instrumental
sensitivity limitations, while the upper bound typically shows a gradual
steepening due to finite-system effects (Pruessner 2012).
Ignoring these effects
leads to power law fits with arbitrary inertial ranges $[x_0, x_2]$,
which affects the accuracy of the determined power law slope and
its uncertainty. Finite-system effects are generally modeled with
an exponential cutoff function (Pruessner 2012), which we can 
combine with the generalized Pareto distribution (Eq.~2),
\begin{equation}
        N(x) dx = n_0 \left( x_0 + x \right)^{-\alpha_x}
        \exp{ \left( - {x \over x_e}\right) } \ dx \ ,
\end{equation}
quantified with the exponential function
$\exp(-x/x_e)$, where $x_e$ is a free variable
when fitting Eq.~(8). Thanks to the large statistics of
Kepler-observed stellar flares, this zone at the upper
bound of the size distribution is not sparsely sampled and
can be accurately modeled. We perform the modified power law
fit functions in Fig.~2, which clearly show an improvement
of our size distribution function (Eq.~8) at the upper end
between $E \approx 10^{36}$ erg and $E \approx 2
\times 10^{37}$ erg, which shows a least-square $\chi^2$
fit improving from $\chi_{diff}=4.5$ (Fig.~1b) to
$\chi_{diff}=1.5$ (Fig.~2b). The inclusion of finite-size
effects changes the power law slope from
$\alpha_E=1.863\pm0.005$ to $\alpha_E=1.817\pm0.005$
(a decrease of $-2.5\%$) for the differential size distribution, and
$\alpha_E=1.877\pm0.005$ to $\alpha_E=1.823\pm0.005$
(a decrease of $-2.9\%$) for the cumulative size distribution
(Figs.~1c and 2c). Thus we conclude that inclusion of finite-system
size effects can change the power law slope by a few percents.

\subsection{	Extreme Events	}

Inspecting the residuals of the power law fits in Fig.~2 we
notice an excess of events at energies of $E \approx
(1 - 4) \times 10^{38}$ erg that cannot be fitted with the exponential
drop-off function, as expected from finite-system size effects 
in the extreme-event zone. These extreme events that deviate
from a standard power law distribution function have also been
dubbed as {\sl ``Dragon-King'' events} by Sornette (2009)
and Sornette and Ouillon (2012), who
suggested that they are generated by a different physical
mechanism. Detections of such extreme event outliers have
occasionally  been noted in astrophysical data sets 
(Aschwanden 2019).

We model the size distribution
with two power law components, where the first power law
distribution includes the exponential function (Eq.~8) with
amplitude $(1-q_{pow})$, while the second power law distribution 
with amplitude $(q_{pow})$ extends all the way to the largest 
event (with an identical power law slope),
\begin{equation}
        N(x) dx = n_0 \left( x_0 + x \right)^{-\alpha_x}
        \left[(1-q_{pow}) \exp{ \left( - {x \over x_e}\right) } 
	+ q_{pow} \right] \ dx \ ,  
\end{equation}
where we define the exponential cutoff energy with $x_e=x_2\ q_{exp}$.
This combinatory definition of the size distribution converges
to the canonical finite-system size distribution as defined
in Eq.~(8), for $q_{pow} \mapsto 0$. We show the best fit of this
model for extreme events in Fig.~3. The $\chi^2$-values amount
to $\chi_{diff}=1.50$ (Fig.~3b) and $\chi_{cum}=0.80$ (Fig.~3c).
The resulting free parameters are $\alpha_{diff}=1.817\pm 0.005$
(Fig.~3b), $\alpha_{cum}=1.821\pm0.005$ (Fig.~3c), $q_{exp}=0.05$,
and $q_{pow}=0.08$ (Fig.~3b). Thus, the extreme events comprise
a fraction of $q_{pow}=8\%$ of all events at high energies of
$x \gapprox x_e$.

A synopsis of the three models is shown in Fig.~4, depicting
the Pareto distribution (P-Model), the finite-system size
effect (F-Model), and the extreme event component (E-Model).
All three models agree with each other in the inertial range
($\gapprox x_0, \lapprox x_3$), but the goodness-of-fit is
in the order of unity for the Models (F) and (E) only. Obviously
the Pareto P-model alone cannot reproduce the high-end tail,
which is reflected in the high values of the best-fit $\chi^2$
values (Fig.~1), while the extreme event model (E) yields the
best consistency with the observed data (Fig.~3). Note that
the best-fitting model (E) has six free parameters
($n_0, a, x_0, x_2, q_{exp}, q_{pow}$).
Such a high degree of precision in the fitting of the
observed size distribution (Fig.~3) is only feasible with
large statistics, such as with the data set of $n_{ev} \gapprox 10^5$ 
stellar flares observed with Kepler.

\subsection{    Estimates of Uncertainties      }

In the binned power law fitting methods we can assert an uncertainty
for the counts per bin from Poisson statistics,
\begin{equation}
        \sigma_{diff,i} = {\sqrt{N_i \Delta x_i} \over \Delta x_i} \ .
\end{equation}
For the cumulative size distribution,
where the difference of events $(N_i-N_{i+1})$ per bin are counted
as independent events only (Aschwanden 2015), the uncertainty is,
\begin{equation}
        \sigma_{cum,i} = \sqrt{(N_i - N_{i+1})} \ .
\end{equation}
Furthermore, the uncertainty $\sigma_\alpha$ of the best-fit power law
slope is estimated to,
\begin{equation}
        \sigma_\alpha = {\alpha \over \sqrt{n_{ev}}} \ ,
\end{equation}
with $n_{ev}$ the total number of events in the entire size distribution
(or in the fitted range), according to Monte-Carlo simulations with
least-square fitting (Aschwanden 2015). A slightly different estimate of
$\sigma_\alpha=(\alpha-1)/\sqrt{n}$ is calculated in Clauset et al.~(2009).

The fitting of any of the 3 models of the differential
occurrence size distribution $N_{diff}^{theo}(x)$ to an
observed (binned) size distribution $N_{diff}^{obs}(x)$,
or of the cumulative size distribution $N_{cum}^{theo}(x)$ to an 
observed size distribution $N_{cum}^{obs}(x)$,
is performed with a standard least-square $\chi^2$ (i.e., 
reduced $\chi^2$) method. The method and numerical Monte-Carlo
simulations of the error estimates are described in more detail
in Aschwanden (2015).

\section{	STATISTICAL RESULTS				}

Here we present
statistics of the means and standard deviations of the reported 
power law slopes in stellar size distributions (Section 4.1),
their dependency on the stellar spectral type (Section 4.2),
their time variability (Section 4.3),
comparisons between stellar and solar power law slopes
(Section 4.4), and the application of self-organized
criticality (SOC) models to stellar flare size distributions
(Section 4.5).

\subsection{	Power Law Slopes of Stellar Flares 	}

We present now a comprehensive compilation of power law slopes
observed from size distributions of stellar flares (with a total of
75 data sets), as listed in Tables 1 and 2. Table 1 includes 33
data sets from the pre-Kepler (1976-2007) era, and 
Table 2 lists 
42 data sets from the Kepler mission (2009-present). 
Pre-Kepler observations of stellar flares have been made
with the UBV, EXOSAT, EUVE, HSP/HST, and XMM-Newton instruments.
When we mention power law slopes of energies
detected in different wavelengths (optical, UV, XUV, 
soft X-rays, or hard X-rays), we should be aware that their
slopes are not assumed to be identical, because they are
produced by different physical mechanisms.

We display the results of the 75 data sets in form of 
scatter plot as a function of the sample size (Fig.~5). 
{\bf All data sets analyzed here and shown in Figs.~5
and 6 are taken from published literature values.}
We divide the 75 data sets into three groups sampled
in different wavelength ranges: The data set with UBV spectroscopy
(Fig.~5a), the data set with XUV fluences, measured with
EXOSAT, EUVE, HSP/HST, XMM-Newton (Fig.~5b), and
the data set OPT with optical luminosities (integrated 
over the flare duration) from Kepler (Fig.~5c). For these 
three groups we find the following mean and standard 
deviations (Fig.~5): 
\begin{equation}
	\alpha_{\rm UBV} =1.80 \pm 0.19 \ ,
\end{equation}
\begin{equation}
	\alpha_{\rm XUV} =2.08 \pm 0.37 \ ,
\end{equation}
\begin{equation}
	\alpha_{\rm OPT} =1.93 \pm 0.40 \ ,
\end{equation}
In order to eliminate small-number samples we select also 
large-size distributions with $n_{ev} > 1000$ events 
(i.e., OPT data set with Kepler data only) and obtain
(Fig.~5c),
\begin{equation}
	\alpha_{\rm OPT}(n_{ev}>1000) =2.09 \pm 0.24 \ ,
\end{equation}
which narrows the distribution of $\alpha$-values from
20\% to 10\%, while their means are consistent within
the uncertainties.

The diagram of Fig.~(5c) illustrates the convergence of
the distribution of $\alpha$-values as a function of the
sample size $n_{ev}$. The small-sample $\alpha$-values spread
over a range of $1.5 \lapprox \alpha \lapprox 3$, while
the large-sample $\alpha$-values (with $n_{ev} > 1000$)
converge to a range of $1.9 \lapprox \alpha \lapprox 2.3$. This
convergence clearly demonstrates that the uncertainty
of the power law slope scales with the sample size,
as predicted by the uncertainty estimate of Eq.~(12)
and shown in Fig.~(5c) (solid curves). 

What is the most accurate value of the fitted 
power law slopes $\alpha$ ?
A compilation of large-number fits can be found in Table 2.
The power law slope of all stellar flare events detected by Kepler 
(n=162,262) is $\alpha=1.817\pm0.005$ (Fig.~3b). 
Inspecting the uncertainties $\sigma_\alpha$ in Fig.~(5c) we notice
that small samples (with $n_{ev} \lapprox 1000$) display uncertainties
that are commensurable with their deviation from a mean value
of $\alpha\approx 2$, while large samples (with $n_{ev} \gapprox 1000$)
display much smaller uncertainties so that they do not overlap
with their common mean value, which indicates that there are
systematic errors that are not accounted for (such as arbitrary
fitting ranges $[x_0, x_2]$ or inaccurate background subtraction).
We do correct for under-sampling of weak stellar fluxes (by
using the thresholded power law model), as well as for 
finite-size effects.  
The cumulative size distribution fit shown in Fig.~(1c) clearly shows 
an anomalous deviation from an ideal power law distribution for 
the largest events in the range of $x \approx 10^{37}$ erg to $4 \times  
10^{38}$ erg, which can be fitted with the finite-system size model (F)
and the extreme event model (E) (Fig.~4).

\subsection{    Power Law Slopes and Stellar Spectral Types }  	 		 

In the study of Yang and Liu (2019), the stellar flare catalog
of the Kepler mission has been sorted according to the
spectral type of the observed stars. This enables us to 
investigate the power law slopes $\alpha$ of the flare size 
(or energy)
distributions for each stellar spectral type (A, F, G, K, M, 
and Giants) separately. Each spectral type may have different 
stellar properties. We show the power law slopes $\alpha$ 
of the various stellar spectral types as a function of the 
sample size in Fig.~6. 
Interestingly, all stellar spectral types 
have power law slopes near $\alpha \approx 2.0$. The
largest data sets (with $n_{ev}>1000$) have the following 
power law slopes: 
$\alpha_G=2.10\pm0.21$ for G-type stars (Fig.~6a), 
$\alpha_F=2.36\pm0.25$ for F-type stars (Fig.~6b), 
$\alpha_M=1.99\pm0.35$ for M-type stars (Fig.~6c),  
$\alpha_K=2.02\pm0.33$ for K-type stars (Fig.~6d), and
$\alpha_{Giants}=1.90\pm0.10$ for Giant stars (Fig.~6f). 
The only exception is $\alpha_A=1.12\pm0.08$ 
for A-type stars (Fig.~6e), 
as it was noted before (Yang and Liu 2019).
This outlier is re-visited in a recent study
and it was found to originate most likely from an 
ill-defined inertial range, while fits with our P, F, and E
models yielded a value of $\alpha_A=1.65\pm0.00$
that is more consistent with the other spectral types
(Aschwanden 2021), and invalidates the claim of a
different physical mechanism made by Yang and Liu (2019).
Thus we ignore this dubious value in further 
analysis.

From the Kepler data shown in 
Fig.~5c it appears that the data sets with
the largest statistics ($n_{ev} > 1000$) with a
power law value of $\alpha_{\rm OPT}=2.09\pm0.24$ 
seem to cluster around the mean value of $\alpha \approx 2.0$.
Given the fact that the same trend holds
for all spectral types (Fig.~6), it is conceivable 
that all stellar types can be explained in terms 
of a single general model that predicts an 
universal value of $\alpha \approx 2$.

\subsection{	Time Variability of Power Law Slopes  	}

We investigate whether
there are evolutionary trends of the slopes $\alpha(t)$ as a 
function of time $t$, since some observations are made up to 30
years apart (during 1970-2000). From the pre-Kepler
data sets (Table 1) we find four cases with substantial
time epochs, all observed in M-dwarf stars (CN Leo,
YX CMi, EV Lac, and AD Leo). We plot the time evolution
of the power law slope $\alpha(t)$ in Fig.~7, which
reveals a highly significant time variation for YZ CMi only,
but this star was observed with different instruments
(i.e., in optical first, and in EUV
later on.) The other cases show temporal variations 
that are commensurable with their uncertainties
$\sigma_\alpha=\alpha/\sqrt{n_{ev}}$ (Eq.~12). Hence
we do not find much evidence for a time evolution of the
power law slopes.

\subsection{ 	Stellar versus Solar Power Laws 	}

For the comparison of stellar versus solar size distributions,
we use the most extensive data set that is available, obtained from 
automated flare detection in {\sl soft X-ray (SXR)} wavelengths, 
using {\sl Geostationary Orbiting Earth Satellite (GOES)} data
from over 37 years, amounting to a total of over $n_{ev} \gapprox$ 300,000 flare 
events (Aschwanden and Freeland 2012). The power law slope 
$\alpha_F$ of the size distribution of GOES 1-4 \ang\ fluxes 
has been measured through three (magnetic) solar activity cycles
(containing 338,661 automatically detected flare events),  
\begin{equation}
	\alpha_{\rm F,GOES} =1.98\pm0.11  
\end{equation}
The soft X-ray fluxes $F_{\rm GOES}$ observed with GOES originate from 
chromospheric evaporation of heated plasma during a flare process. 
According to the Neupert effect (Dennis and Zarro 1993), 
the soft X-ray flux is essentially the time integral of the hard 
X-ray flux $f_{\rm HXR}(t)$, which represents 
the heating rate of the soft X-ray emitting plasma,
\begin{equation}
	P_{\rm SXR} = \int f_{\rm HXR}(t) \  dt \approx F_{\rm HXR} \ T 
	\approx E_{\rm HXR} \ .
\end{equation}  
This relationship implies identical power law slopes of
$\alpha_{\rm P,SXR} = \alpha_{\rm E,HXR}$, which can be used to
constrain the free parameters ($S, \beta, \gamma, D_S$) 
of the FD-SOC model. 

The power law slopes obtained from GOES soft X-ray fluxes
$\alpha_{\rm F,GOES} = 1.98 \pm 0.11$ (Eq.~17)
are similar to those of the UBV data set (Fig.~5a; 
$\alpha_{\rm UVB}=1.80\pm0.19$), 
with those in XUV wavelengths (Fig.~5b;
$\alpha_{\rm XUV}=2.08\pm0.37$, and with those of the Kepler data sets, 
with $\alpha_{\rm OPT} =1.93\pm0.40$ for all 40 optical 
datasets (Fig.~5c), which converges to $\alpha_{\rm OPT}=
2.09\pm 0.24$ (Fig.~5c) for large data sets with $n_{ev} > 1000$. 
The self-consistency of power law slopes
$\alpha_F \approx 2.0$ corroborates the proportionality 
($\gamma \approx 1$) between (optically-thin) fluxes 
and (fractal) flare volumes (Eq.~24), as assumed in our
fractal-diffusive SOC model. 

How typical is our Sun among other stars of the same spectral
type G? In Fig.~(6a) we depict the power law slopes of 13
datasets that comprise stellar flare statistics of solar-like 
stars, groups that are referred to as
G-types (Yang and Liu 2019; Wu et al.~2015), G5-stars
(Maehara et al.~2012; Shibayama et al.~2013; 
Aschwanden 2015), slowly rotating G5-stars (Maehara et al.~2012;
Shibayama et al.~2013), and G dwarfs (Audard et al.~1999).
The diagram in Fig.~(6a) reveals that all of these solar-like
data sets of G type are consistent with a power law slope
value of $\alpha_F \approx 2$, within the stated uncertainties
(as listed in Table 2). The largest of these G-type data sets
from Kepler (with n=55,259 flare events) displays a value of 
$\alpha_G=1.96\pm0.04$ (Yang and Liu 2019), which appears to
converge to an asymptotic value of the power law slope 
$\alpha_F \approx 2.0$
(Fig.~6a). Thus, we can conclude that both solar and stellar 
flares have approximately similar power law slopes in their size 
distributions, but we have to be aware that solar flares are
observed in hard X-rays and soft X-rays, while stellar flares 
are observed in optical wavelengths here.

\subsection{	Self-Organized Criticality and Stellar Flares }	

The concept of {\sl Self-Organized Criticality (SOC)} has
been introduced by Bak et al.~(1987, 1988) and has been
applied to a large number of nonlinear processes in the
universe (see textbooks by Aschwanden 2011; Pruessner 2012).
In a nutshell, we can characterize SOC models in terms of
a critical state of a nonlinear energy dissipation system
that is slowly and continuously driven towards a critical
value of a system-wide instability threshold, producing
scale-free, fractal-diffusive, and intermittent avalanches
with powerlaw-like size distributions (as defined in
Aschwanden 2014; 2015). Avalanches in such nonlinear
systems exhibit an exponential growth phase and a random-like
duration, resulting in power law-like size distributions
(called scale-free), while the spatio-temporal
behavior can exhibit fractal geometry, diffusive
transport, and intermittent time structures.
Classical paradigms for SOC models
are sand avalanches, earthquakes, forest fires, as well as
a number of astrophysical processes, such as solar and
stellar flares. Our prime question here is whether a SOC
model can predict the exact value of the power law slope
for stellar flare phenomena, or whether observations of
stellar flares can confirm and constrain the power law
slopes predicted by SOC models. A secondary question in this study
is whether flares on active stars have identical size distributions
as solar flares, and thus are likely to be produced by the
same physical mechanism. 

We follow the concept of a generalized SOC system as defined
in Aschwanden (2014). The assumptions made therein consist
of the following 6 physical relationships:

(1) The scale-free probability conjecture, which simply
says that the number of avalanches $N(L)$ with size $L$ is 
reciprocal to the length scale $L$ (normalized by $L_0$),
with Euclidean dimension $D$, 
\begin{equation}
	N(L)\ dL \propto \left( {L \over L_0} \right)^{-D}\ dL  \ ,
\end{equation} 
(2) the Euclidean flare volume $V$ is fractal and has a fractal 
(Hausdorff) dimension $D_d$,
\begin{equation}
	V \propto L^{D_d} \ ,
\end{equation}
(3) the mean fractal dimension $D_d$ is given by the geometric mean
of the smallest ($D_{min} \approx 1$) and largest possible
fractal dimension $D_{max} \approx D$,
\begin{equation}
	D_d \approx {D_{d,min} + D_{d,max} \over 2}
	= {(1 + D) \over 2} \ ,
\end{equation}
(4) the evolution of an avalanche as a function
of time duration $T$ grows according to classical diffusion,
\begin{equation}
	L \propto T^{\beta/2} \ \mapsto \ T \propto L^{2/\beta} \ , 
\end{equation}
(5) the observed flux $F$ scales with the avalanche volume $V$
with a power law exponent $\gamma$,
\begin{equation}
	F \propto V^\gamma \ ,
\end{equation}
(6) the peak flux is given by avalanche propagation with
maximal fractal dimension $(D_{max})$, and the avalanche energy
$E$ is given approximately by the product of the mean flux $F$
and the avalanche duration $T$,
\begin{equation}
	E = F \ T \ .
\end{equation}
From these 6 assumptions we can derive straightforwardly
the predicted power law slopes of the size distributions
of a SOC system. 
Since all relationships given in Eq.~(19-24) are expressed in terms
of the variable $L$, the power law indices $\alpha_x$ for the
size distribution of the parameters $x=A, V, T, F, P, E$
can be derived by substitution of variables,
i.e., $N(x) dx =\ N[x(L)]\ |dx/dL]\ dL \propto x^{-\alpha_x}$,
yielding the power law indixes $\alpha_x$ (Aschwanden 2019),
\begin{eqnarray}
                 \alpha_L &=& S \approx 3 \\
                 \alpha_A &=& 1 + (S - 1)/D_2 \approx 7/3 \\
                 \alpha_V &=& 1 + (S - 1)/D_3 \approx 2 \\
                 \alpha_T &=& 1 + (S - 1) \beta/2 \approx 2 \\
                 \alpha_F &=& 1 + (S - 1)/(\gamma\ D_S) \approx 2 \\
                 \alpha_P &=& 1 + (S - 1)/(\gamma\ S) \approx 5/3 \\
                 \alpha_E &=& 1 + (S - 1)/(\gamma\ D_S + 2/\beta ) \approx 3/2 \ ,
\end{eqnarray}
where we included the dimensionality of the Euclidean
geometry (S), the flare area (A), and the peak flux (P) also.
The approximative numerical values are obtained by inserting the
default parameters $S=3, \beta=1, \gamma=1, D_2=1.5, D_3=2$,
based on  observations
(see Aschwanden 2014). In other words, our basic assumptions
for SOC avalanches boil down to the three hypotheses of a 
3-D Euclidean space ($S=3$), classical diffusion ($\beta=1$), 
and fractal geometry ($D_{min}=1, D_{max}=3$).

\subsection{ 	Low-Cadence Kepler Data	}

Since Kepler supposedly measures the (time-integrated) 
fluences $E \propto F_{\rm SXR}$ (Eq.~18), 
we expect a power law slope of $\alpha_E=1.5$ 
for the fluence (Eq.~31), rather than the observed value
of $\alpha_{\rm F,GOES} \approx 2.0$ (Eq.~17).
This apparent discrepancy can be reconciled in
the case of flare detection with insufficient 
cadence, i.e., $\Delta T \ge T$,
in which case the fluence relationship $E = F\ T$ 
(Eq.~24) degenerates to proportionality,
\begin{equation}
	E_K \approx F\ \Delta T \propto F \ ,
\end{equation}
where the subscript ``K'' refers to the Kepler-specific 
low-cadence mode.  From the proportionality
between the energy $E_K$ and the mean flux $F$ we expect then 
an identical power law slope, 
\begin{equation}
	\alpha_{EK} = \alpha_F = 2.0 \ ,
\end{equation}
since the time cadence $\Delta T$ is a constant.
Indeed, long-cadence data from Kepler are processed in
time intervals ($2 \Delta T$ = 1 hour) that are substantially
longer than the typical duration of solar flares 
(e.g., $T=25\pm30$ min; Aschwanden 2020). 
Flare durations on young brown stars observed in the
Kepler low-cadence (1 min) mode were found over a range
of $T=0.17-8.6$ min (Gizis at al.~2017).
However, the vast majority of stars were observed 
using the long, 30-minute cadence mode (Davenport 2016).
Each continuous time segment was required to be at least 
two days in duration, and any segment less than two days 
in duration was discarded from analysis (Davenport 2016).
Filter widths with lower limits of 6 hrs and upper limits
of 24 hrs were used in the analysis of Yang and Liu (2019).
If stellar flare durations are similar to solar
flare durations, most of the stellar flares would not be
resolved in low-cadence data from Kepler.

\subsection{ Background Subtraction Effect } 

The emitted radiation from flares in astrophysical objects
contains almost always a flare-unrelated background flux
in every wavelength, which could bias the measurement of
the power law slope in a size distribution (Appendix A). 
This slope
is severely modified in the lowest decade of a size
histogram, while it does not affect the slope of large
sizes. The background subtraction effect has been simulated
in detail (Aschwanden 2015), which we corroborate with
additioal data given in Table 3. We sorted the power law
measurements with subtracted background in the first
column of Table 3, which yields a mean value of,
\begin{equation}
	\alpha_E^{BG} = 1.57 \pm 0.19 \ ,
\end{equation}
which is close to the theoretical prediction of the
standard SOC model (i.e., $\alpha_E=1.5$). There is 
also some statistics available from mostly GOES data
without (or insufficient) background subtraction
(Table 3, second column),
\begin{equation}
	\alpha_E^{NBG} = 2.20 \pm 0.22 \ ,
\end{equation}
which clearly corroborates the trend that the neglect
of flare-unrelated background subtraction causes a mean
bias of $(\alpha_E^{NBG}-\alpha^{BG}) \lapprox 0.6 $ in
the power law slope $\alpha_E$. At this point we do not
know how accurately the flare-unrelated background
has been treated in the 75 stellar flare data sets
discussed here (Tables 1 and 2). However, the fact
that many power law slopes of the fluence size 
distribution exhibit values around $\alpha_E \approx 2$
(Fig.~5), lets us suspect that some of the high values
could have been generated by inaccurate background
subtractions. 

While the empirical result of
a power law slope of $\alpha_F \approx 2$ has been established 
for stellar flare events recently (Yang and Liu 2019), we are
adding here a theoretical explanation for this finding that is
predicted by SOC models. Taking the filtering of short flare
durations in Kepler data into account, we find that the SOC
model predicts the proportionalities $E_K \propto F$ (Eq.~32) 
and $F \propto V$ (Eq.~23), which implies the identities 
$\alpha_{EK} = \alpha_F = \alpha_V = 2$ and agrees with
the observations generally (Fig.~5).

\section{	Discussion	}

\subsection{	Flare Frequency and Stellar Rotation 	}

The rotation period of a star is expected to be related
to the star's age, with younger stars rotating more rapidly,
at least after their evolution of several 100 Myr on the
main sequence; for younger stars, the rotation periods
are widely distributed, depending on the ``initial''
rotation period (Johnstone et al.~2020).
However it was found that the maximum energy of the flare 
is not correlated with the stellar rotation period, but 
the data suggest that superflares (with energies 
$E \ge 10^{33}$ erg) occur more frequently on rapidly 
rotating stars (Maehara et al.~2012). The frequency of 
superflares on slowly rotating stars was found to be
smaller than that of all G-type dwarfs, and the frequency
of superflares on hot G-type dwarfs ($5600 < T_{eff} < 6000$ K)
is smaller than for cool G-type dwarfs ($5100 < T_{eff} < 5500$ K).  
In general, stars with shorter rotation periods tend to have 
larger power law slope values $\alpha$. 
It has been confirmed that spots on flare stars are generally larger 
than those on non-flaring stars and that flare stars rotate
significantly faster than non-flaring stars (Balona 2015).
For stars in this sample with previously measured rotation 
periods, the total relative flare luminosity (Appendix B)
has been compared to 
the Rossby number. A tentative detection of flare activity 
saturation for low-mass stars with rapid rotation below a 
Rossby number of $\approx 0.03$ has been found (Davenport 2016). 
The rotation distribution of flare stars shows that about 70\% 
of flare stars rotate with periods shorter than 10 days and 
the rate approaches 95\% at 30 days (Yang and Liu 2019). 

Comparing stellar flares
from different spectral types involves
often a strong detection bias. For instance,
the bolometric luminosity of an M dwarf is
very small compared to a G star, and thus
smaller flares can be detected on M stars than
on G stars for the same absolute sensitivity
(Balona 2015). But it is also true that G stars
can produce larger flares in absolute terms 
than M dwarfs (Johnstone et al. 2020).

\subsection{	Nanoflares Versus Noise 	}

A lingering question in stellar observations is the
unknown composite structure of apparently quiescent emission,
especially from dM stars. In other words: Does quiescent 
emission entirely consist of Poisson noise, or is it made
up of microflares (or nanoflares) that are part of a continuous 
size distribution extending with the same power law slope
to large flares (Collura et al.~1988)? One would expect
that this question can simply be settled by the characterization
of their size distribution, which should be different for  
exponential-like or power law-like functions. 
Early observations of K and M dwarf stars with EXOSAT 
(Schmitt and Rosso 1988; Pallavicini et al. 1990; 
McGale et al.~1995) did not have sufficient sensitivity 
to detect coherent microflaring activity.
The detection of the smallest flaring structures strongly
depends on the detection threshold. 
Examination of EUV time profiles from
RS CVn binaries reveals both, small-scale stochastic
variability, as well as small flares (Osten and Brown 1999).
More sensitive observations with Einstein-IPC
did report continuous variability
(Ambruster et al.~1987). Nearly continuous
flaring in both X-rays and correlated fluxes
in the U band were detected with EUVE 
(Audard et al. 2000; Guedel et al. 2002, 2004).
In some samples of Kepler data, 
for instance, only flares have been included with total energies
of $E \ge 5 \times 10^{34}$ erg (Shibayama et al. 2013). 
Those energies derived from Kepler observations are still orders
of magnitudes more powerful than the largest solar flares
$E \approx 10^{30}-10^{33}$, while solar nanoflares have been
detected down to $E \gapprox 10^{24}$ erg (Krucker and Benz 1998;
Parnell and Jupp 2000; Aschwanden and Parnell 2002). 

The power law index of stellar size distributions 
($\alpha_{EK} \approx 2.0$) is 
larger than for typical solar flares in hard X-rays
($\alpha_E \approx 1.5$), 
including small-scale solar events. If the power law 
continues to energies of moderate solar flares, then the total 
energy emitted by the ensemble of all flares may suffice to 
explain all of the observed flaring and “quiescent” X-ray 
emissions of flare stars. A considerable portion, if not all, 
of the energy required to heat their coronae could thus be 
provided by flares (Audard et al.~1999; 2000).
The main result of this paper  and of Yang and Liu (2019)
is that stellar flares show a similar power law distribution slope
of $\alpha \approx 2$ as solar flares 
do, and thus both solar and stellar flares
(including microflares and nanoflares)
are consistent with a common physical 
mechanism for flare heating and the
coupled coronal heating.

\subsection{	SOC in Stellar Flares	}

What does it mean when we say that stellar flares exhibit 
self-organized criticality ? If stellar flares would occur
by pure random processes, their size distribution would fit
a Poissonian or Gaussian function. In contrast, the fact that
stellar flares are consistent with power law functions strongly
supports the evolution of nonlinear (exponential-growing)
energy dissipation processes, triggered by local 
fluctuations that exceed a system-wide threshold (Aschwanden 2011).     
The statistics of physical parameters in such nonlinear energy 
dissipation processes can be expressed with volumetric
scaling laws, characterized by the scale-free probability
(Eq.~19), the (spatial) fractal dimension (Eqs.~20, 21), 
classical diffusion (Eq.~22), the flux-volume scaling (Eq.~23),
and the fluence scaling (Eqs.~24, 32). SOC models predict
cross-correlations and power law size distributions with
specific slopes in energy scaling laws, which can 
be tested with the observed parameters (Aschwanden 2020). 
The energy scaling laws depend strongly on the dimensionality
and geometry of dissipation events. The spatio-temporal
behavior of flare events may be governed by classical diffusion
(Eq.~22). The fractal structure of a flaring region may quantify
the self-similar geometry of the filling factor in a post-flare 
arcade. For stellar flares, we cannot test any spatial geometry
due to the remoteness of the observed stars, 
so we cannot determine geometric slopes directly
($\alpha_L, \alpha_A, \alpha_V$), but we have 
four observed power law slopes that can be compared with 
predictions i.e., $\alpha_P=5/3=1.67$ for the peak flux, 
$\alpha_F=2$ for the mean flux, 
$\alpha_T=2$ for the flare durations, and
$\alpha_E=3/2=1.5$ for the fluences. 
Since most of the quoted 20 publications provide the
energy $E$ only, we cannot fully test the SOC model
with stellar flare data. Tests of the SOC model with solar
flare data (Aschwanden et al.~2015, 2020), however, reveal 
encouraging results, although
a number of improvements in the data analysis could be
carried out (including data truncation, instrumental
sensitivity threshold, preflare background subtraction,
fine-size effects, etc.) (Aschwanden 2020), as
demonstrated in Figs.~1-3.

\section{	Conclusions				}	

We study the size distributions of stellar flare events,
compiled from 20 publications, which contain a variety of 
75 data sets from different stars and stellar spectral types,
characterized by their power law slope $\alpha$ of their
size distribution $N(x) \propto x^{-\alpha_x}$. The largest
data set comes from the Kepler mission, which produced a
stellar flare catalog with 162,262 events, for which we
perform high-precision modeling of its size distribution.
We summarize the conclusions in order of the 5 initial
questions.

\begin{enumerate}

\item{\underbar{What are the values and uncertainties of power law
slopes in size distributions of stellar flares ?}
Size distributions of nonlinear events have generally been
fitted with a power law function that extends over an arbitrary
inertial range $[x_1, x_2]$, which causes large uncertainties
in the slope value if the inertial range extends over a small
range of 1-2 decades. This problem is ameliorated for the large
stellar flare data set of the Kepler mission, which contains
$n_{ev}\gapprox 10^5$ events and covers an inertial range of
$\lapprox 4$ decades in energy. A second source of uncertainties
is the roll-over of the size distribution at small events
(near $x_1$) due to incomplete sampling, data truncation, 
background subtraction, and instrumental sensitivity limitations,
which can be adequately modeled with a thresholded power law
or Pareto distribution. A third source of uncertainties is the
neglect of an exponential drop-off at the upper end of the
size distribution due to finite-system size effects, which
can be modeled with an exponential function. A fourth uncertainty
is the occasional presence of extreme events at the largest
energies (near $x_2$) beyond the finite-system size effects,
which are also called {\sl Dragon-King events} and indicate
a different (secondary) physical mechanism than the primary
mechanism that produces the primary power law events.
Taking all these effects into account, we find a best-fit
power law slope value of $\alpha=1.817\pm0.005$ for the
entire Kepler data set. Averaging all 75 stellar data sets 
yields a range of $\alpha_{\rm OPT}=1.93\pm0.40$, while the largest 
11 data sets with $n_{ev} >1000$ events per set yield a 
range of $\alpha_{\rm OPT}=2.09\pm0.24$.}

\item{\underbar{Do power law slopes vary with time ?}
We identified four M-dwarf stars with multiple observations
of stellar flare size distributions during different times.
We observe significant variations of the power law slope
in time for YX CMi only, but they were measured with 
different instruments.
Two other stars (AD Leo, CN Leo) show marginal variation.
The variation in the power law slope $\alpha(t)$ could
possible indicate stellar magnetic cycles of order $5-20$ years.}

\item{\underbar{Do power law slopes of stellar flares depend on the
spectral stellar type ?}
We find no significant deviations from the average power law slope
$\alpha \approx 2$ for the spectral stellar types G, F, M, K, Giants.
There is an outlier of $\alpha \approx 1.1$ for spectral type A in
the study of Yang and Liu (2019), but automated fitting with our (P, F, E) 
models yield a more reliable value of $\alpha_{E}=1.65$ for all
3 models (Aschwanden 2021),
which implies that the outlier $\alpha=1.1$ represents an erroneous value due 
to small-number statistics and an inadequately chosen fitting range.}

\item{\underbar{Are stellar flare size distributions compatible with 
solar flares ?}
Comparing the fluence size distributions of solar flares 
observed with GOES in soft X-ray wavelengths 
$(\alpha_{\rm F,GOES}=1.98\pm0.11)$ (Eq.~17) 
with those from Kepler observed in optical wavelengths
$(\alpha_{\rm F,OPT}=2.09\pm0.24)$ (Eq.~16),
we find a good agreement between these two power law slopes,
one measured in solar flares, the other in stellar flares.
Futhermore, the Neupert effect observed in solar flares
(Dennis and Zarro 1993) predicts a power law slope of 
$\alpha_{\rm F,SXR} \propto \alpha_{\rm E,HXR}= 2.0$. 
Strictly speaking, a proper
comparison between solar and stellar flares would require
white-light data from both instruments, which are sparsely
available for solar data.
The energy range of all Kepler flare events amounts to 
$E \approx 10^{34} - 4 \times 10^{38}$ erg, while solar 
flares cover a range of $E \approx 10^{24} - 10^{33}$ erg.}

\item{\underbar{Are stellar flares consistent with self-organized
criticality (SOC) models ?}
Standard SOC models predict power law slopes of $\alpha_F=2.0$ 
for the mean flux, $\alpha_V=2.0$ for the flare volume,
and $\alpha_E=1.5$ for the flare fluence, which changes
to a value of $\alpha_{EK}=2.0$ when the flare durations
are not resolved or uncorrelated, which is the case in the analysis 
of low-cadence Kepler data cited here. Therefore we can conclude
that white-light emissivity is proportional to the flare 
volume. Since the predicted values $\alpha_F = \alpha_V 
= \alpha_{EK} = 2.0$ overlap with the observed 
range of stellar flares, $\alpha_{\rm OPT}=2.09\pm0.24$, 
stellar size distributions are consistent with SOC models.} 

\end{enumerate}

The Kepler mission has provided us an enormous wealth
of stellar data, since stellar flares have been detected
in a total of 3420 flare stars, while we had previously
flare data from a single star (our Sun) only. 
Kepler detected a total of 162,262 stellar
flares, which is a compatible with the largest
solar flare data set, obtained from GOES
(with 338,661 flare events).
On the energetic side, the largest
solar flare (with an energy of $E \lapprox 10^{33}$ erg)
is still below the smallest flare detected with Kepler,
with $E \gapprox 10^{34}$ erg. Therefore, solar and
stellar flare data sets are complementary in energy,
but have very similar size distributions with power
law slopes of $\alpha \approx 2$. Adding solar and
stellar flares together would yield a combined 
size distribution
that extends over 14 orders of magnitude (from $10^{24}$ 
erg to $10^{38}$ erg)! It needs to be shown that the
same physical flare mechanism holds for such a large
range of energies. 

\clearpage

\section*{APPENDIX A: Background Subtraction in Solar Flares	}

One of the largest systematic uncertainties results from the
preflare background subtraction, because the preflare flux
is often not specified in solar flare catalogs. As long as
the power law fits in size distribtutions of fluxes or
fluences are applied in inertial ranges that are sufficiently
above the background flux level, there is no problem with
fitting power law slopes, but it interoduces a systematic
bias, typically from a slope of $\alpha_E \approx 1.5$
in background-subtracted data to $\alpha_E \approx 2.0$
for size distributions sampled without background subtraction.
In the case of no background subtraction, the bias is always 
directed towards a steeper value, because the correction
implies a relatively larger shift of small flares towards 
the right-hand side of a size distribution than for large
flares. This effect has been simulated in detail in
Fig.~3 of Aschwanden (2015).

On the observational side we find the following information
on the preflare background in hard X-rays and gamma-rays: 
Early on, background subtraction was obtained manually
between preflare background and peak flux (Drake 1971);
HXRBS/SMM has a mean flare-unrelated
background of $\approx 40$ cts s$^{-1}$ in the $>25$ keV
flux (Dennis et al.~1991; Crosby et al.~1993);
ISEE-3 detects a mean background of $\approx 24$ cts s$^{-1}$
at $>50$ keV, or 50 cts s$^{-1}$ at $>30$ keV (Bromund et al.~1995);  
background-subtracted data of gamma-ray flares are used from
WATCH/GRANAT and GRS/SMM (Perez Enriquez and Mirochnishenko 1999);
A global solar background with a 3-sigma threshold was set to 
detect microflares with RHESSI (Christe et al.~2008).

In soft X-ray wavelengths, reliable preflare background values 
can be obtained from images, such as from Yohkoh/SXT (Shimizu 1995).
Flare detection using time profiles of GOES without background
subtraction leads to a substantial bias for too steep power law
slopes, which is the case in a number of studies: 
(Veronig et al.~2002a, 2002b; Yashiro et al.~2006).
The preflare background has also been estimated from the absolute 
minimum of a time series encompassing each nanoflare 
(Krucker and Benz 2002; Benz and Krucker 2002), 
which yields a lower limit of the
background fluence, and thus produces a similar bias
as no background-subtraction. A similar search for nanoflares
in TRACE data (Parnell and Jupp 2000) exhibits a sharp peak at
the low end of the fluence size distribution, which is typical 
for under-estimated background subtraction (Aschwanden 2015). 
A more reliable background has been determined from imaging
data, by mapping out the excess emission measure (associated 
with each nanoflare) in TRACE 171 and 195 filter data 
(Aschwanden et al.~2000b; Aschwanden and Parnell 2000a,b),
or using SOHO/EIT imaging data (Uritsky et al.~2007, 2013).

In Table 3 we give a compilation of published solar flare 
size distributions of the fluence, $E = F\ T$, tabulated
separately for studies that employed methods with
preflare background subtraction (Table 3, first column),
and without background subtraction (Table 3, second column).
The systematic background subtraction bias can be clearly
be seen when the two groups are averaged separately.
While the background-subtracted samples yield a mean
fluence value of $\alpha_E = 1.57\pm0.11$ (or a median
value of 1.54), which agrees with the theoretically
predicted value of $\alpha_E=1.5$ according to the
fractal-diffusive SOC model (Aschwanden 2015), while
those statistics without preflare background subtraction
yield a systematically higher value of $\alpha_E=
2.10\pm0.27$.  Ideally, the preflare backgrounds should be
subtracted for each flare event individually, but
if this information is not available, at least a
mean background value should be corrected.

\section*{APPENDIX B: The Kepler Flare Energy Definition	}

There are two measures to quantify the energy of a stellar
flare: (i) the (background-subtracted) peak (flux) or amplitude
of a stellar luminosity time profile $f_{flare}(t)$ in a 
given wavelength range $\lambda\pm \Delta\lambda$,
which has a mean of $F_{flare}=<f_{flare(t)}>$, 
$$
	P_{flare} = max[f_{flare}(t)] \ ,
	\eqno(B1) 
$$
and (ii) the time-integrated luminosity (intensity, or energy) 
$E_{flare}$,
$$
	E_{flare} = \int f_{flare}(t) \ dt \approx
	F_{flare} \ T_{flare} \ ,
	\eqno(B2) 
$$
also called fluence or total energy, where $T_{flare}$
is the flare duration. 

For the energy measurements of the Kepler instrument, 
it is assumed that the spectrum of white-light flares can 
be described by a black-body radiation with an effective 
temperature of $T_{flare}=10^4$ K and flare area $a_{flare}(t)$
(Shibayama et al.~2013),
$$
	f_{flare}(t)= \sigma_{SB} T_{flare}^4(t) \ a_{flare}(t) \ ,
	\eqno(B3) 
$$
with $\sigma_{SB}$ the Stefan-Boltzmann constant.
The flare area $a_{flare}(t)$ is estimated from the observed
luminosity of the star $(f_{star}')$, the flare $(f_{flare}')$,
and the flare amplitude of the light curve $f_{flare}(t)$ of 
Kepler after detrending, 
$$
	f_{star}' = \int R_{\lambda}
	B_{\lambda(T_{eff})} \ d\lambda \ \cdot \pi R_{star}^2, \ 
	\eqno(B4) 
$$
$$
	f_{flare}'(t) = \int R_{\lambda}
	B_{\lambda(T_{flare}(t))} d\lambda \cdot a_{flare}(t)\ , 
	\eqno(B5) 
$$
$$
	a_{flare}(t) = \pi R^2 
	\left( {f_{flare}' \over f_{star}'} \right)
	{\int R_\lambda B_{\lambda(T_{eff})} d \lambda \over 
	 \int R_\lambda B_{\lambda(T_{flare}(t))} d \lambda} \ ,  
	\eqno(B6) 
$$
where $B_{\lambda(T)}$ is the Planck function, and $R_{\lambda}$
is the response function of the Kepler instrument.

If we compare the total energy of a stellar flare observed 
by Kepler with solar flares, the corresponding quantity is the
bolometric energy of white-light emission. However, white-light
emission in solar flares is observed for the largest flares only,
produced by $\gapprox 5$ keV electrons and ions that precipitate
into the deeper chromosphere and produce long-lived excess
ionization in the heated chromosphere to enhance free-free and
free-bound continuum emission, visible in broadened Balmer
and Paschen lines (Hudson 1972). The ratio of bolometric
energies $E_{bol}$ to the thermal energy $E_{th}$ of the flare
plasma was found to be commensurable, i.e., $E_{bol}/E_{th}
=1.14\pm0.05$ (Kretzschmar 2011; Aschwanden et al.~2015, 2017). 

\vskip1cm
{\sl Acknowledgements:}
Part of the work was supported by NASA contract NNG04EA00C of the
SDO/AIA instrument and the NASA STEREO mission under NRL contract
N00173-02-C-2035.

\clearpage


\section*{ References }  

\def\ref#1{\par\noindent\hangindent1cm {#1}} 

\ref{Ambruster, C.W., Sciortino, S., and Golub, L. 1987,
	{\sl Rapid, low-level X-ray variability in active 
	late-type dwarfs}, ApJS 65, 273}
\ref{Arzner, K. and G\"udel, M. 2004,
        {\sl Are coronae of magnetically active stars heated by flares?
        III. Analytical distribution of superposed flares},
        ApJ 602, 363}
\ref{Arzner, K., Guedel, M., Briggs, K., Telleschi, A., and Audard, M. 2007,
 	{\sl Statistics of superimposed flares in the Taurus molecular cloud},
 	AA 468, 477}
\ref{Aschwanden, M.J., Nightingale, R., Tarbell, T., and Wolfson, C,J. 2000a,
 	{\sl Time variability of the quiet Sun observed with TRACE. 
	I. Instrumental effects, event detection, and discrimination of EUV nanoflares},
 	ApJ 535, 1027}
\ref{Aschwanden, M.J., Tarbell, T., Nightingale, R., Schrijver, C.J., 
	Title, A., Kankelborg, C.C., Martens, P.C.H., and Warren, H.P. 2000b,
 	{\sl Time variability of the quiet Sun observed with TRACE. 
	II. Physical parameters, temperature evolution, and energetics of EUV nanoflares}
 	ApJ 535, 1047}
\ref{Aschwanden, M.J. and Parnell, C.E. 2002,
 	{\sl Nanoflare statistics from first principles: fractal geometry and 
	temperature synthesis},
 	ApJ 572, 1048}
\ref{Aschwanden, M.J. 2011,
 	{\sl Self-Organized Criticality in Astrophysics. The Statistics of 
	Nonlinear Processes in the Universe}, 
 	ISBN 978-3-642-15000-5, Springer: New York}
\ref{Aschwanden, M.J., and Freeland, S.L. 2012,
	{\sl Automated solar flare statistics in soft X-rays over 37 years
	of GOES observations: The invariance of self-organized criticality
	during three solar cycles},
	ApJ 754:112} 
\ref{Aschwanden, M.J. 2014,
	{\sl A macroscopic description of a generalized self-organized
	criticality system: Astrophysical applications},
	ApJ 782:54}
\ref{Aschwanden, M.J. 2015,
	{\sl Thresholded powerlaw size distributions of instabilities 
	in astrophysics}, ApJ 814, 19}
\ref{Aschwanden, M.J., Caspi, A., Cohen, C.M.S., Holman, G.D., 
	Jing, J., Kretzschmar, M., Kontar  E.P., McTiernan,J.M., et al. 2017,
 	{\sl Global energetics of solar flares: V. Energy closure},
 	ApJ 836:17}
\ref{Aschwanden, M.J. 2019,
	{\sl Self-organized criticality in solar and stellar flares:
	Are extreme events scale-free ?}
	ApJ 880, 105}
\ref{Aschwanden, M.J. 2020,
	{\sl Global energetics of solar flares. XII. Energy scaling laws},
	ApJ 903:23}
\ref{Aschwanden, M.J. 2021,
	{\sl Finite system-size effects in self-organized criticality systems},
	ApJ, (subm.)}
\ref{Audard, M., G\"udel, M., and Guinan, E.F. 1999,
        {\sl Implications from extreme-ultraviolet observations for
        coronal heating of active stars},
        ApJ 513, L53}
\ref{Audard, M., Guedel, M., Drake, J.J., and Kashyap, V.L. 2000,
 	{\sl EUV flare activity in late-type stars},
 	ApJ 541, 396}
\ref{Bak, P., Tang, C., and Wiesenfeld, K. 1987,
        {\sl Self-organized criticality - An explanation of 1/f noise},
        Physical Review Lett. 59/27, 381}
\ref{Bak, P., Tang, C., and Wiesenfeld, K. 1988,
        {\sl Self-organized criticality},
        Physical Rev. A 38/1, 364}
\ref{Benz, A.O. and S. Krucker 2002,
 	{\sl Energy distribution of micro-events in the quiet solar corona}
 	ApJ 568, 413}
\ref{Borucki, W.J., Koch, D., Basri, G. et al. 2010,
	{\sl Kepler planet-detection mission: Introduction and
	first results}, Science 327, 977}
\ref{Bromund, K.R., McTiernan, J.M., and Kane, S.R. 1995,
 	{\sl Statistical studies of ISEE3/ICE observations of 
	impulsive hard X-ray solar flares}
 	ApJ 455, 733}
\ref{Christe, S., Hannah, I.G., Krucker, S., McTiernan, J., and Lin, R.P. 2008,
 	{\sl RHESSI microflare statistics. I. Flare-finding and 
	frequency distribution},
 	ApJ 677, 1385}
\ref{Clauset, A., Shalizi, C.R., and Newman, M.E.J. 2009,
 	{\sl Power-law distributions in empirical data
 	SIAM Review 51/4, 661}
\ref{Collura, A., Pasquini, L., and Schmitt, J.H.M.M. 1988,
	{\sl Time variability in the X-ray emission of dM stars 
	observed by EXOSAT},
 	AA 205, 197}
\ref{Crosby, N.B., Aschwanden, M.J., and Dennis, B.R. 1993,
	{\sl Frequency distributions and correlations of
	solar flare X-ray flare parameters},
	SoPh 143, 275.}
\ref{Davenport, J.R. 2016,
	{\sl The Kepler Catalog of stellar flares},
	ApJ 829:23}
\ref{Dennis, B.R., Orwig, L.E., Kennard G.S., Labow, G.J.,
	Schwartz, R.A., Shaver, A.R., and Tolbert, A.K. 1991,
	{\sl The complete Hard X-Ray Burst Spectrometer
	event list, 1980-1989}
	NASA Technical Memorandum 4332}
\ref{Dennis, B.R. and Zarro, D.M. 1993,
	{\sl The Neupert effect: What can it tell us about
	the impulsive and gradual phases of solar flares},
	SoPh 146, 177}
\ref{Drake, J.F. 1971,
 	{\sl Characteristics of soft solar X-ray bursts},
 	SoPh 16, 152}
\ref{Gizis, J.E., Paudel, R.R., Mullan, D.,Schidt, S.J.,
	Burgasser, A.J., and WIlliamer, P.K.G. 2017,
	{\sl K2 Ultracool dwarf survey. II. The white light flare rate
	of young brown dwarfs}
	ApJ 845:33} 
\ref{G\"udel, M., Audard, M., Skinner, S.L., and Horvath, M.I. 2002, 
	{\sl X-ray evidence for flare density variations and
	continual chromospheric evaporation in Proxima Centauri},
	ApJ 580, L73}
\ref{G\"udel, M., Audard, M., Kashyap, V.L., and Guinan, E.F. 2003,
        {\sl Are coronae of magnetically active stars heated by flares?
        II. Extreme Ultraviolet and X-ray flare statistics and the
        differential emission measure distribution},
        ApJ 582, 423}
\ref{G\"udel, M., Audard, M., Reale, F., Skinner, S.L., and Linsky, J.L. 
	et al. 2004, 
 	{\sl Flares from small to large: X-ray spectroscopy of 
	Proxima Centauri with XMM-Newton}
	AA 416, 713}
\ref{Hosking, J.M.R. and Wallis, J.R. 1987,
	{\sl Parameter and quantile estimation for the 
	generalized Pareto distribution},
	Technometrics 29(3), 339.}
\ref{Hudson, H.S. 1972, {\sl }
 	{\sl Thick target processes and white light flares},
	SoPh 24, 414}
\ref{Johnstone, C.P., Bartel, M., and G\"udel, M. 2020,
	{\sl The active lives of stars: a complete description of rotation
	and XUV evolution of F, G, K, and M dwarfs?}
	AA (in press)}
\ref{Kashyap, V.L., Drake, J.J., G\"udel, M., and Audard, M. 2002,
        {\sl Flare heating in stellar coronae},
        ApJ 580, 1118}
\ref{Kretzschmar, M. 2011, 
	{\sl The Sun as a star: observations of white-light flares}, 
	AA 530, A84}
\ref{Krucker, S. and Benz, A.O. 1998,
 	{\sl Energy distribution of heating processes in 
	the quiet solar corona},
 	ApJ 501, L213}
\ref{Lacy, C.H., Moffett, J.J., and Evans, D.S. 1976, 
 	{\sl UV Ceti stars: statistical analysis of observational data},
 	ApJSS 30, 85}
\ref{Lomax, K.S. 1954, 
	{\sl Business Failures; Another example of the 
	analysis of failure data}, 
	J. Am. Stat. Assoc. 49, 847}
\ref{Maehara, H., Shibayama, T., Notsu, Y., Nagao, T., Kusaba, S.,
	Honda, S., Nogami, D., and Shibata, K. 2012,
	{\sl Superflares on solar-like stars},
	Nature 485, 478}
\ref{McGale, P.A., Pye, J.P., Barber, C.R., and Page, C.G. 1995
	Temporal behaviour of sources in the ROSAT 
	extreme-ultraviolet all-sky survey},
	MNRAS 275, 1232}
\ref{Moffett, J.J. 1974, {\sl UV Ceti flare stars observational data},
 	ApJSS 273, 29, 1}
\ref{Moffett, J.J. and Bopp, B.W. 1976, 
 	{\sl UV Ceti stars: statistical analysis of observational data},
 	ApJSS 31, 61} 
\ref{Osten, R.A. and Brown, A. 1999,
	{\sl Extreme Ultraviolet Explorer (EUVE) photometry of RS 
	Canum Venaticorum (RS CVn) systems: four flaring megaseconds},
 	ApJ 515, 746}
\ref{Pallavicini, R., Tagliaferri, G., and Stella, L. 1990,
 	{\sl X-ray emission from solar neighbourhood flare stars: 
	a comprehensive survey of EXOSAT results},
 	AA 228, 403}
\ref{Parnell, C.E. and Jupp, P.E. 2000,
 	{\sl Statistical analysis of the energy distribution of 
	nanoflares in the quiet Sun},
 	ApJ 529, 554}
\ref{Perez-Enriquez, R. and Miroshnichenko, L.I. 1999,
 	{\sl Frequency distributions of solar gamma ray events 
	related and not related with SPEs in 1980-1995},
 	SoPh 188, 169}
\ref{Pruessner, G. 2012,
 	{\sl Self-organised criticality. Theory, models and 
	characterisation},
 	ISBN 9780521853354, Cambridge University Press: Cambridge}
\ref{Robinson, R.D., Carpenter, K.G., and Percival, J.W. 1999,
        {\sl A search for microflaring activity on dMe flare stars.
        II. Observations of YZ Canis Minoris},
        ApJ 516, 916}
\ref{Schmitt J.H.M.M. and Rosso, C. 1988,
	{\sl EXOSAT observations of M dwarf stars in the solar
	neighborhood},
	AA 191, 99}
\ref{Shibayama, T., Maehara, H., Notsu, S., Nosu, Y.,
	Nagao, T., Honda, S., Ishi, T.T., Nogami, D., and Shibata, K. 2013,
	{\sl Superflares on solar-type stars observed with Kepler. I.
	Statistical properties of superflares},
	ApJSS 209:5}
\ref{Shimizu, T. 1995,
 	{\sl Energetics and occurrence rate of active-region transient 
	brightenings and implications for the heating of the 
	active-region corona}, 
	PASJ 47, 251}
\ref{Sornette, D. 2009,
	{\sl Dragon-King, black swans and the prediction of crises}, 
	J. Terraspace Science and Engineering, 2, 1}
\ref{Sornette, D. and Ouillon, G. 2012,
	{\sl Dragon-Kings: Mechanisms, statistical methods
	and empirical evidence},
	EPJST 205, 1} 
\ref{Stelzer, B., Flaccomio, E., Briggs, K., Micela, G., Scelsi, L,
        Audard, M., Pillitteri, I., and G\"udel, M. 2007,
        {\sl A statistical analysis of X-ray variability in pre-main
        sequence objects of the Taurus molecular cloud},
        AA 468, 463}
\ref{Uritsky, V.M., Paczuski, M., Davila, J.M., and Jones, S.I. 2007,
 	{\sl Coexistence of Self-Organized Criticality and Intermittent 
	Turbulence in the Solar Corona},
 	Phys.Rev.Lett. 99, 025001}
\ref{Uritsky, V.M., Davila, J.M., Ofman, L., and Coyner, A.J. 2013,
 	{\sl Stochastic Coupling of Solar Photosphere and Corona},
	ApJ 769, 62}
\ref{\bf Van Doorsselaere, T., Hoda, S., and Deboscher J. 2017,
	{\sl Stellar flares observed in long-cadence data from the
	Kepler mission},
	ApJSS 232:26.}
\ref{Veronig, A.M., Temmer, M., Hanslmeier, A., Otruba, W., and Messerotti, M. 2002a,
 	{\sl Temporal aspects and frequency distributions of solar soft X-ray flares},
 	AA 382, 1070}
\ref{Veronig, A.M., Temmer, M., and Hanslmeier, A. 2002b,
	{\sl Frequency distributions of solar flares},
	Hvar Observatory Bulletin, 26, 7} 
\ref{Wu, C.J., Ip, W.H., and Huang, L.C. 2015,
	{\sl A study of variability in the frequency distributions
	of the superflares of G-type stars observed by the Kepler
	mission}, ApJ 798:92}
\ref{Yang, H. and Liu, J. 2019,
	{\sl The flare catalog and the flare activity in the
	Kepler mission},
	ApJSS 241:1}
\ref{Yashiro, S., Akiyama, S., Gopalswamy, N., and Howard, R.A. 2006,
 	{\sl Different Power-Law Indices in the Frequency Distributions 
	of Flares with and without Coronal Mass Ejections},
 	ApJ 650, L143} 
\clearpage


\begin{table}
\begin{center}
\normalsize
\caption{The power law slopes of the frequency distributions of
energies observed in stellar flares: $\alpha$ for the
differential frequency distribution, and $\alpha = \beta+1$ 
for the cumulative size distributions.}
\begin{tabular}{lllllll}
\hline
Star	   & Spectal	& Instrument & Number    & Differential  & Cumulative     & Reference\\ 
	   & type       &	     & of flares & slope	 & slope plus 1   &          \\
           & 		&            & or events & $\alpha$      & $\beta+1$  &          \\
\hline
\hline
CN Leo      & dM6.5 Ve 	& UBV  & 111  & 1.99$\pm$0.19 & $1.99\pm0.12$ & Lacy et al. 1976 \\ 
UV Cet	    & M6.0e	& UBV  & 107  & 1.98$\pm$0.19 & $1.98\pm0.14$ & Lacy et al. 1976 \\
Wolf 424 AB & M5.5	& UBV  & 11   & 1.81$\pm$0.54 & $1.81\pm0.18$ & Lacy et al. 1976 \\
YZ CMi	    & M5 V  	& UBV  & 62   & 1.71$\pm$0.22 & $1.71\pm0.08$ & Lacy et al. 1976 \\
EQ Peg      & M4 Ve  	& UBV  & 58   & 2.00$\pm$0.26 & $2.00\pm0.14$ & Lacy et al. 1976 \\
EV Lac	    & M3.5      & UBV  & 22   & 1.69$\pm$0.36 & $1.69\pm0.11$ & Lacy et al. 1976 \\
AD Leo      & M3.5eV    & UBV  & 9    & 1.82$\pm$0.61 & $1.82\pm0.27$ & Lacy et al. 1976 \\
YY Gem      & dM1e      & UBV  & 6    & 1.43$\pm$0.58 & $1.43\pm0.11$ & Lacy et al. 1976 \\
M dwarfs    & dM        & EXOSAT  & 13 & 1.52$\pm$0.42 &               & Collura et al. 1988$^*)$\\
M dwarfs    & dM        & EXOSAT  & 20 & 1.70$\pm$0.10 &               & Pallavicini et al. 1990\\
RS CVns     & G         & EUVE    & 30 & 1.60$\pm$0.32 &               & Osten and Brown 1999\\
YZ CMi      & M5 V      & HSP/HST & 54 & 2.25$\pm$0.31 & $2.25\pm0.10$ & Robinson et al. 1999\\
G dwarfs    & G         & EUVE    & 25 & 2.15$\pm$0.15 &               & Audard et al. 1999\\
HD 2726     & F2 V      & EUVE    & 15 & 2.43$\pm$1.07 & 2.61$\pm$0.38 & Audard et al. 2000\\
47 Cas      & G0-5 V    & EUVE    & 12 & 2.62$\pm$1.85 & 2.19$\pm$0.34 & Audard et al. 2000\\
EK Dra      & G1.5 V    & EUVE    & 15 & 1.78$\pm$0.62 & 2.08$\pm$0.34 & Audard et al. 2000\\
$\kappa$ Cet 1994 & G5 V& EUVE    & 5  & 2.55$\pm$0.31 & 2.18$\pm$0.89 & Audard et al. 2000\\
$\kappa$ Cet 1995 & G5 V& EUVE    & 10 & 2.45$\pm$1.08 & 2.29$\pm$0.51 & Audard et al. 2000\\
AB Dor      & K1 V      & EUVE    & 16 & 1.76$\pm$0.74 & 1.88$\pm$0.26 & Audard et al. 2000\\
$\epsilon$ Eri & K2 V   & EUVE    & 12 & 2.38$\pm$1.50 & 2.40$\pm$0.81 & Audard et al. 2000\\
GJ 411      & M2 V      & EUVE    & 15 & 1.57$\pm$0.57 & 1.63$\pm$0.29 & Audard et al. 2000\\
AD Leo      & M3.5 V    & EUVE    & 12 & 1.65$\pm$1.17 & 2.02$\pm$0.28 & Audard et al. 2000\\
EV Lac      & M4.5 V    & EUVE    & 12 & 1.75$\pm$1.18 & 1.76$\pm$0.33 & Audard et al. 2000\\
CN Leo 1994 & M6.5 Ve   & EUVE    & 13 & 2.24$\pm$0.63 & 2.21$\pm$0.30 & Audard et al. 2000\\
CN Leo 1995 & M6.5 Ve   & EUVE    & 15 & 1.59$\pm$0.86 & 1.46$\pm$0.39 & Audard et al. 2000\\
FK Aqr      & dM2e      & EUVE    & 65 & 2.60$\pm0.34$ &               & Kashyap et al. 2002\\
V1054 Oph   & M3 Ve     & EUVE    & 85 & 2.74$\pm0.35$ &               & Kashyap et al. 2002\\
AD Leo      & M3.5 V    & EUVE    & 80 & 2.17$\pm0.03$ &               & Kashyap et al. 2002\\
AD Leo      & M3.5 V    & EUVE    & 75 & 2.32$\pm0.11$ &               & Kashyap et al. 2002\\
AD Leo      & M3.5 V    & EUVE    & ?  & 2.25$\pm0.25$ &               & G\"udel et al. 2003\\
AD Leo      & M3.5 V    & EUVE    & ?  & 2.30$\pm0.10$ &               & Arzner\& G\"udel 2004\\
HD 31305    & M, K      & XMM-Newton & ?  & 2.20$\pm0.30$ &               & Arzner et al. 2007\\
TMC         & M, K      & XMM-Newton & ?  & 2.40$\pm0.50$ &               & Stelzer et al. 2007\\
\hline
\end{tabular}
\end{center}
\par $^*$) The flare fluxes from multiple stars were co-added 
            without distance correction in Collura et al.~(1988).
\end{table}


\begin{table}
\begin{center}
\normalsize
\caption{The power law slopes of the frequency distributions of
energies observed in stellar flares with Kepler: $\alpha$ for the
differential frequency distribution, and $\alpha = \beta+1$ 
for the cumulative size distributions.}
\begin{tabular}{llllll}
\hline
Star	   & Instrument & Number    & Differential  & Cumulative     & Reference       \\
	   &		& of flares & slope	    & slope plus 1   &                 \\
           &            & and events& $\alpha$      & $\beta+1$  &                 \\
\hline
\hline
G5-stars        &Kepler         &  365           & 2.30$\pm$0.30  &               &Maehara et al. 2012  \\
G5-stars slow   &Kepler         &  101           & 2.00$\pm$0.20  &               &Maehara et al. 2012  \\
G5-stars        &Kepler         & 1547           & 2.20$\pm$0.06  &		 &Shibayama et al. 2013\\
G5-stars slow   &Kepler         &  397           & 2.00$\pm$0.10  &               &Shibayama et al. 2013\\
G5-stars        &Kepler         & 1538           & 2.43$\pm$0.08 & 2.42$\pm$0.06 &Aschwanden  2015     \\
G-type stars    &Kepler         & 4494           & 2.04$\pm$0.17 &               &Wu et al. 2015 \\
K-M,A-F stars   &Kepler         &  209           & 1.69$\pm$0.16 & 1.71$\pm$0.12 &Balona 2015, Aschwanden 2015\\
KID3557532      &Kepler         &  196           & 2.11$\pm$0.19 &               &Wu et al. (2015) \\
KID6034120      &Kepler         &   45           & 3.12$\pm$0.60 & 3.17$\pm$0.48 &Aschwanden 2015 \\
KID6697041      &Kepler         &   37           & 1.83$\pm$0.37 & 1.51$\pm$0.25 &Aschwanden 2015 \\
KID6865416      &Kepler         &  147           & 1.77$\pm$0.10 &               &Wu et al. 2015 \\
KID75264976     &Kepler         &   40           & 1.92$\pm$0.34 & 1.98$\pm$0.32 &Aschwanden 2015 \\
KID7532880      &Kepler         &  159           & 1.90$\pm$0.16 &               &Wu et al. 2015 \\
KID8074287      &Kepler         &  160           & 1.87$\pm$0.10 &               &Wu et al. 2015 \\
KID8479655      &Kepler         &   39           & 1.45$\pm$0.23 & 1.47$\pm$0.24 &Aschwanden 2015 \\
KID8547383      &Kepler         &   40           & 3.41$\pm$0.68 & 2.58$\pm$0.41 &Aschwanden 2015 \\
KID9653110      &Kepler         &  158           & 1.64$\pm$0.07 &               &Wu et al. 2015 \\
KID10422252     &Kepler         &  177           & 1.75$\pm$0.08 &               &Wu et al. 2015 \\
KID10422252     &Kepler         &   57           & 2.99$\pm$0.58 & 2.78$\pm$0.37 &Aschwanden 2015 \\
KID10745663     &Kepler         &  137           & 1.63$\pm$0.10 &               &Wu et al. 2015 \\
KID11551430     &Kepler         &  202           & 1.59$\pm$0.06 &               &Wu et al. 2015 \\
Stellar flares  &Kepler         &  208           & 1.68$\pm$0.12 &               &Aschwanden 2019\\
Bolometric flares&Kepler        & 1537           & 2.55$\pm$0.07 &               &Aschwanden 2019\\
KID10000490     &Kepler         &  241           &     & 1.55$\pm$0.10 &Davenport 2016 \\ 
KID10001145     &Kepler         &  271           &     & 2.40$\pm$0.15 &Davenport 2016 \\ 
KID10001154     &Kepler         &  119           &     & 1.56$\pm$0.14 &Davenport 2016 \\ 
KID10001167     &Kepler         &  147           &     & 1.41$\pm$0.12 &Davenport 2016 \\ 
KID10007792     &Kepler         &  225           &     & 1.52$\pm$0.10 &Davenport 2016 \\ 
KID10002897     &Kepler         &  155           &     & 1.28$\pm$0.10 &Davenport 2016 \\ 
KID10004510     &Kepler         &  142           &     & 1.43$\pm$0.12 &Davenport 2016 \\ 
KID10004660     &Kepler         &  135           &     & 1.83$\pm$0.24 &Davenport 2016 \\ 
KID10005966     &Kepler         &  175           &     & 1.79$\pm$0.14 &Davenport 2016 \\ 
KID10006158     &Kepler         &  279           &     & 1.85$\pm$0.11 &Davenport 2016 \\ 
A-type		&Kepler		&  583		 & 1.12$\pm$0.08 & $1.65\pm0.07$ &Yang \& Liu 2019, Aschwanden 2021\\
F-type		&Kepler		&  8869	 	 & 2.11$\pm$0.09 & $1.85\pm0.09$ &Yang \& Liu 2019, Aschwanden 2021\\
G-type		&Kepler		&  55259	 & 1.96$\pm$0.04 & $1.79\pm0.06$ &Yang \& Liu 2019, Aschwanden 2021 \\
K-type		&Kepler		&  47112	 & 1.78$\pm$0.02 & $1.80\pm0.20$ &Yang \& Liu 2019, Aschwanden 2021 \\
M-type		&Kepler		&  50439	 & 2.13$\pm$0.05 & $1.84\pm0.02$ &Yang \& Liu 2019, Aschwanden 2021 \\
Giants 		&Kepler		&  6496	 	 & 1.90$\pm$0.10 & $2.00\pm0.34$ &Yang \& Liu 2019, Aschwanden 2021 \\
All		&Kepler         &  162262        &               & 1.823$\pm$0.007 &This work \\
\hline
\end{tabular}
\end{center}
\end{table}

\begin{table}
\begin{center}
\normalsize
\caption{The power law slopes of the frequency distributions of
solar flare fluences $\alpha_E$, calculated with preflare
background subtraction (first column) or without any background
subtraction (second column).}
\begin{tabular}{llrlll}
\hline
Powerlaw   & Powerlaw   & Number of & Wavelength or     & Instrument    & Reference\\
slope of   & slope of	& events    & energy threshold 	&		& 	\\
fluence $\alpha_E$ & fluence $\alpha_E$	& $n_{ev}$ & 	& 		&	\\
background & no background &	&			&		&	\\
subtracted & subtracted	&	&		 	&		&	\\
\hline
1.53$\pm0.02$ &		& 7045	& $>$25 keV 	& HXRBS/SMM & Crosby et al.~1993 \\
1.71$\pm0.04$ &		& 1008	& $>$25 keV	& HXRBS/SMM & Crosby et al.~1993 \\
1.68$\pm0.07$ &		&  545	& $>$25 keV	& HXRBS/SMM & Crosby et al.~1993 \\
1.67$\pm0.03$ &		& 3874	& $>$25 keV	& HXRBS/SMM & Crosby et al.~1993 \\
1.69$\pm$0.02 &	 	& 4356  & $>$30 keV	& ISEE-3    & Bromund et al.~1995\\
1.80$\pm$0.01 &         &  110  & $>$100 keV    & PHEBUS, GRS/SMM & Perez Enriquez et al.~(1999)\\
1.38$\pm$0.01 &         &  110  & $>$75 keV     & PHEBUS, GRS/SMM & Perez Enriquez et al.~(1999)\\
1.39$\pm$0.01 &         &  185  & $>$300 keV    & PHEBUS, GRS/SMM & Perez Enriquez et al.~(1999)\\
1.30$\pm$0.01 &         &   67  & $>$300 keV    & PHEBUS, GRS/SMM & Perez Enriquez et al.~(1999)\\
1.50$\pm$0.03 &         &   67  & $>$511 keV    & PHEBUS, GRS/SMM & Perez Enriquez et al.~(1999)\\
1.39$\pm$0.02 &         &   67  & $>$2223 keV   & PHEBUS, GRS/SMM & Perez Enriquez et al.~(1999)\\
1.31$\pm$0.02 &         &   67  & $>$1000 keV   & PHEBUS, GRS/SMM & Perez Enriquez et al.~(1999)\\
1.39$\pm$0.01 &         &  134  & $>$1000 keV   & PHEBUS, GRS/SMM & Perez Enriquez et al.~(1999)\\
1.70$\pm$0.10 &         & 4241  & $>$12 keV     & RHESSI    & Christe et al.~(2008)\\
1.44	      &         & 4028  & 2-12 \ang\    & Explorer  & Drake (1971) \\
1.55$\pm$0.05 &         & 5008  & 1265 \ang\    & Yohkoh/SXT& Shimizu (1995)\\
 	      & 2.03$\pm$0.09 & & 0.5-4 \ang\    & GOES      & Veronig et al.~(2002a)\\
 	      & 1.89$\pm$0.10 & & 0.5-4 \ang\    & GOES      & Veronig et al.~(2002b)\\
 	      & 2.01$\pm$0.03 & & 0.5-4 \ang\    & GOES      & Yashiro et al.~2006\\
 	      & 2.3-2.6 &       & 171, 195 \ang\ & SOHO/EIT  & Krucker and Benz (1998)\\
 	      & 2.0-2.6	&	& 171, 195 \ang\ & TRACE,SOHO/EIT & Parnell and Jupp (2000)\\
1.79$\pm$0.08 &         &       & 171, 195 \ang\ & TRACE     & Aschwanden et al.~(2002a,b)\\
              & 2.31-2.59 &     & 171, 195 \ang\ & SOHO/EIT  & Benz and Krucker (2002)\\
              & 2.04-2.52 &     & 171, 195 \ang\ & SOHO/EIT  & Benz and Krucker (2002)\\
2.06$\pm$2.10 &          &      & 171 \ang\      & TRACE     & Aschwanden and Parnell (2002)\\
1.70$\pm$0.17 &          &      & 195 \ang\      & TRACE     & Aschwanden and Parnell (2002)\\
1.41$\pm$0.09 &          &      & AlMag          & Yohkoh/SXT& Aschwanden and Parnell (2002)\\
1.54$\pm$0.03 &          & 	& 171, 195, AlMg & Yohkoh/SXT& Aschwanden and Parnell (2002)\\
1.66-1.70     &          & 	& EUV            & SOHO/EIT  & Uritsky et al.~(2007)\\
1.54$\pm$0.03 &          & 	& 171, 195       & SOHO/EIT  & Uritsky et al.~(2013)\\
\hline 
1.57$\pm$0.19 & 2.20$\pm$0.22 & &                &           & Mean observed value  \\
1.54          & 2.28          & &                &           & Median observed value\\
1.50 	      &               & &	 	 &           & SOC model prediction \\
\hline
\end{tabular}
\end{center}
\end{table}


\begin{figure}
\centerline{\includegraphics[width=0.9\textwidth]{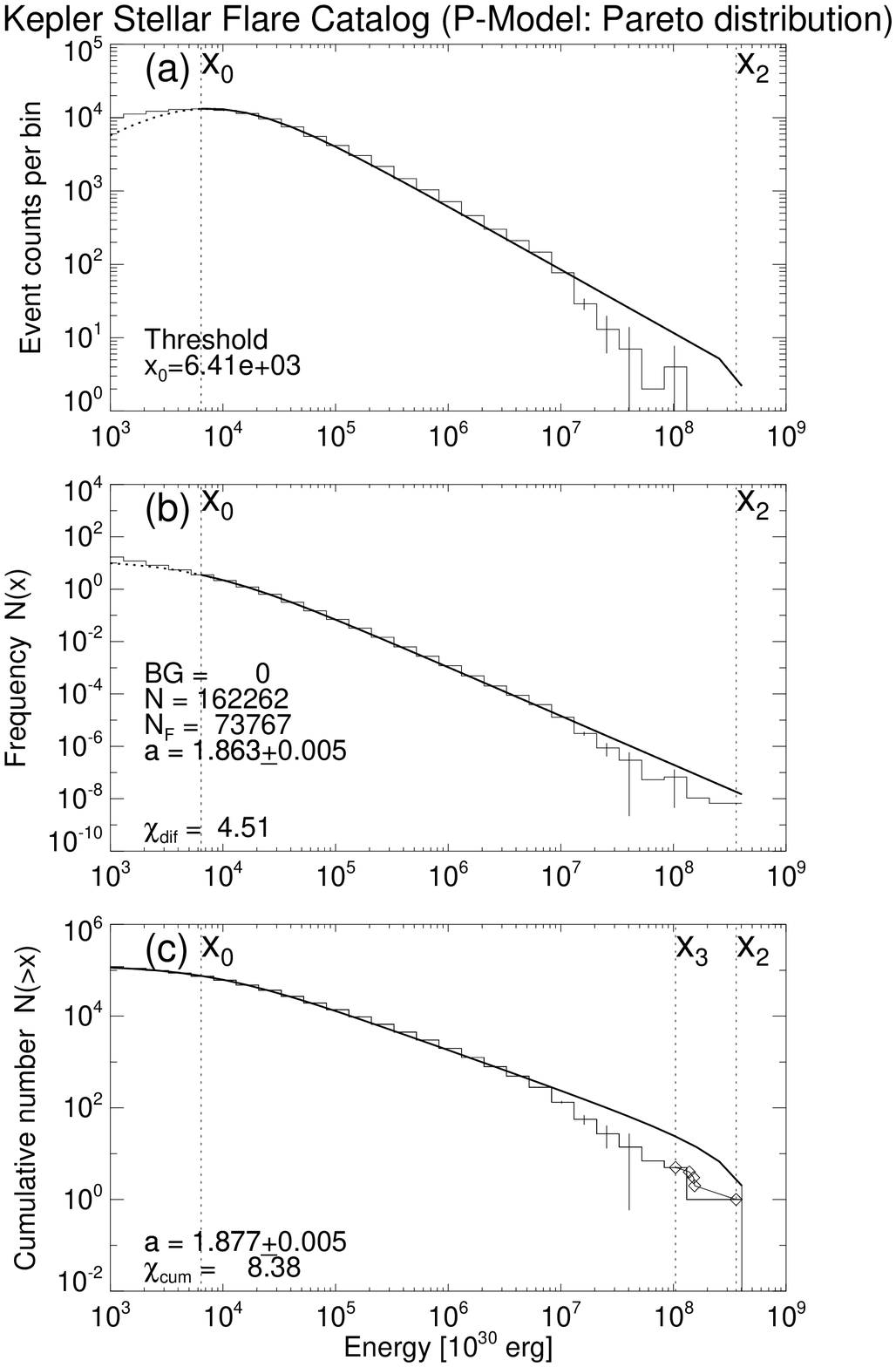}}
\caption{(a) Flare event counts in (logarithmic) histogram bins; 
(b) differential occurrence frequency distribution; 
(c) cumulative occurrence frequency distribution; 
and the rank-order bins of the 5 largest events (diamonds). 
The entire Kepler stellar flare catalog contains 162,262 events.
A thresholded power law (Pareto distribution) function is shown
(with solid curves), which represents an approximate model of
the undersampling of small events.} 
\end{figure}

\begin{figure}
\centerline{\includegraphics[width=0.9\textwidth]{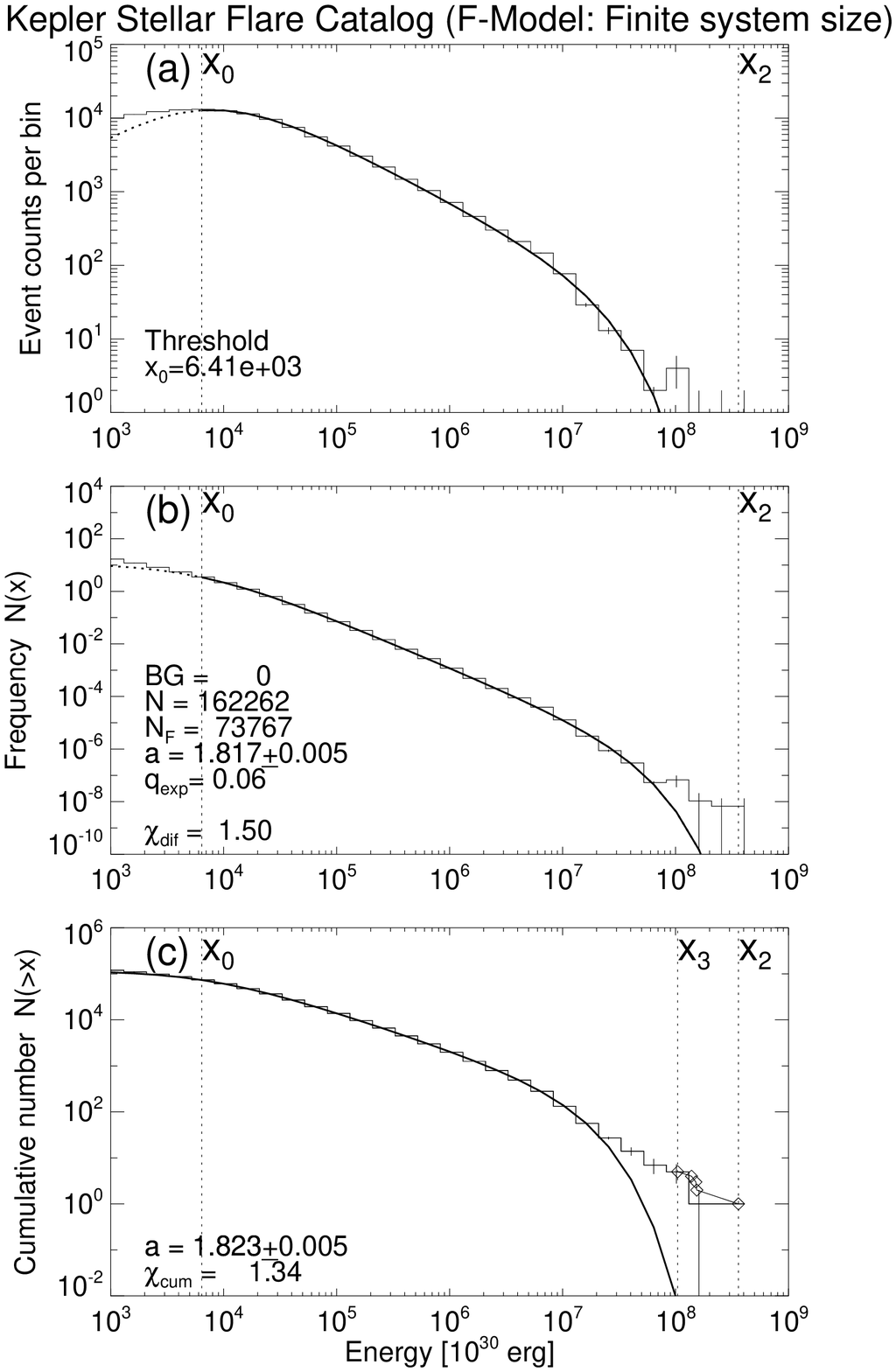}}
\caption{Stellar flare events from Kepler data base 
are fitted with an exponential drop-off function that models 
finite system size effects. Otherwise similar representation 
as in Fig.~1.}  
\end{figure}

\begin{figure}
\centerline{\includegraphics[width=0.9\textwidth]{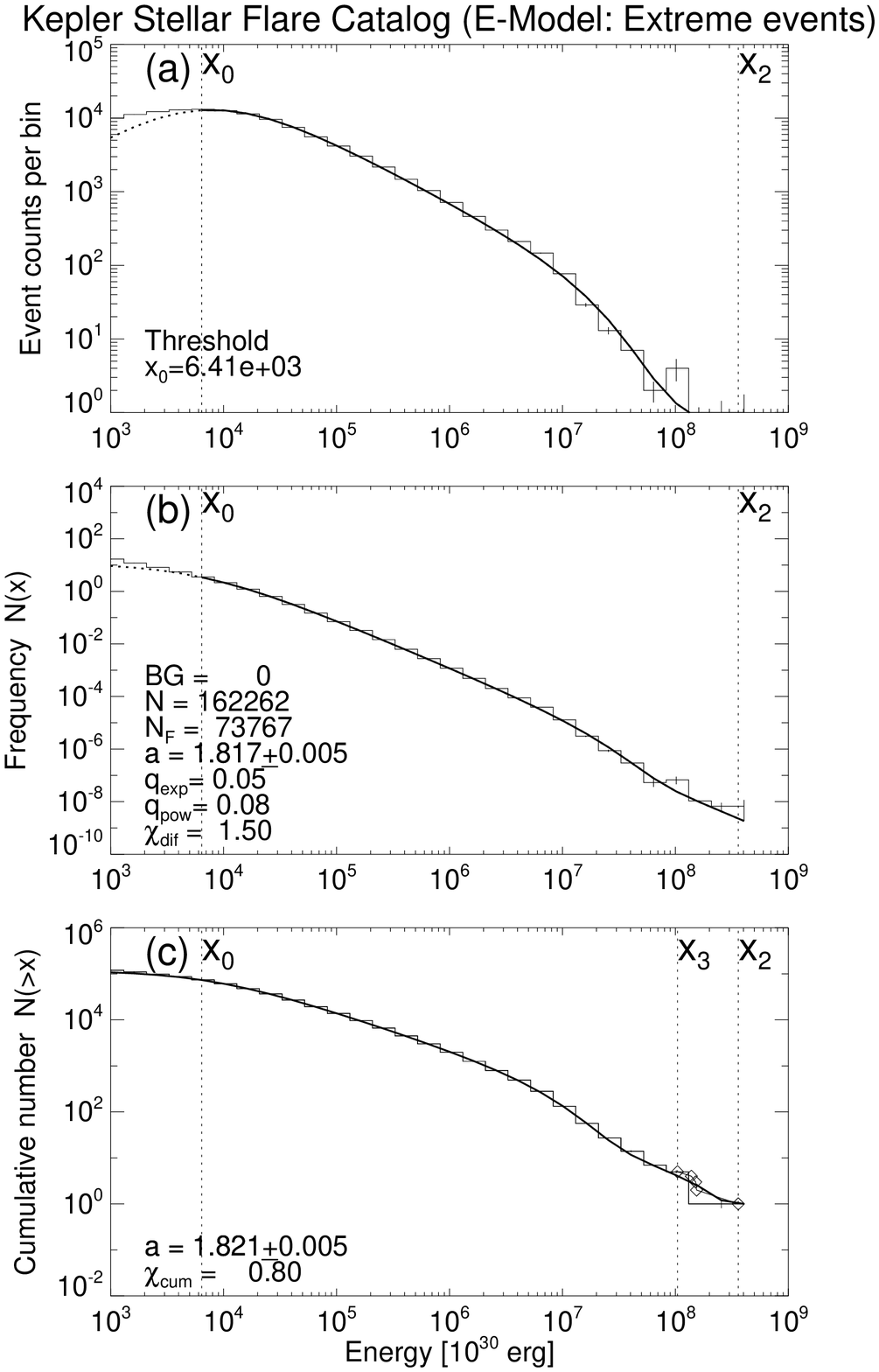}}
\caption{Stellar flare extreme events from Kepler data base 
are fitted with a secondary power law component. Otherwise
similar representation as in Figs.~1 and 2.}  
\end{figure}

\begin{figure}
\centerline{\includegraphics[width=0.9\textwidth]{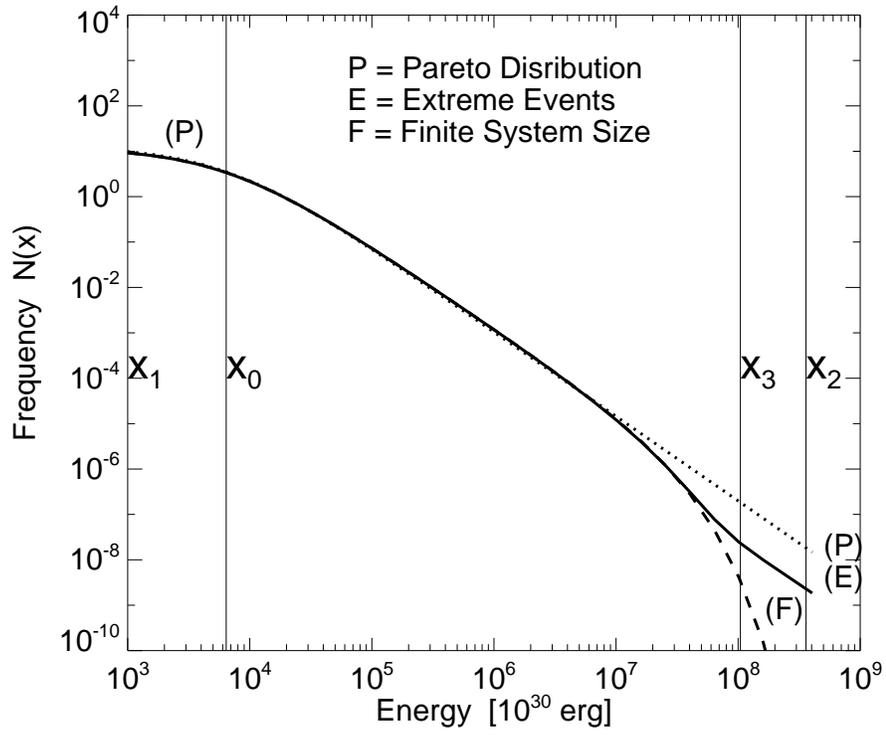}}
\caption{Synopsis of three power law models: 
(P) = Pareto distribution $[x_1, x_0]$,
(E) = Extreme events $[x_3, x_2$], and 
(F) = Finite system size model $[\lapprox x_3]$.
The inertial range covers $[\gapprox x_0, \lapprox x_3$].}
\end{figure}

\begin{figure}
\centerline{\includegraphics[width=1.0\textwidth]{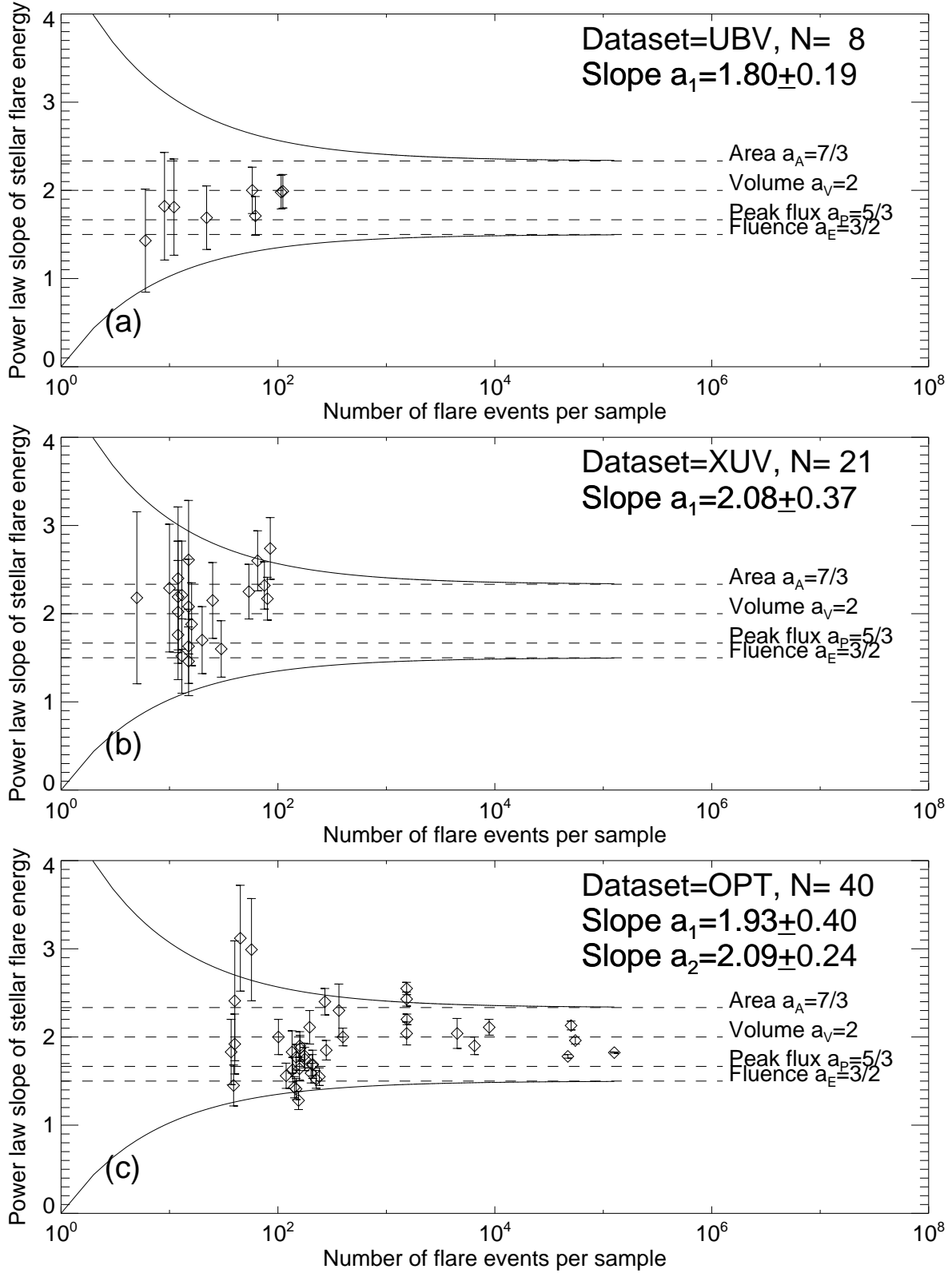}}
\caption{The power law slopes ($\alpha$) of flare energy distributions
as a function of the number of events per sample are shown separately
for three instrument groups or wavelength ranges: UBV (a), 
XUV [EXOSAT, EUV, HSP/HST, XMM-Newton] (b), and OPT [Kepler] (c).
The solid curves bracket the range of statistical uncertainties. 
The dashed horizontal lines indicate the predicted slopes of 
four SOC parameters:
for the size distribution of 
the flare volume ($\alpha_V=2$), 
the flare area ($\alpha_A=7/3=2.33$),
the peak flux ($\alpha_P=5/3$),
and the time-integrated flare energy or fluence ($\alpha_E=3/2=1.5$).
The slope $a_1$ includes all data sets, the slope $a_2$ samples
with large statistics only ($n_{ev} \ge 1000$).}
\end{figure}

\begin{figure}
\centerline{\includegraphics[width=0.8\textwidth]{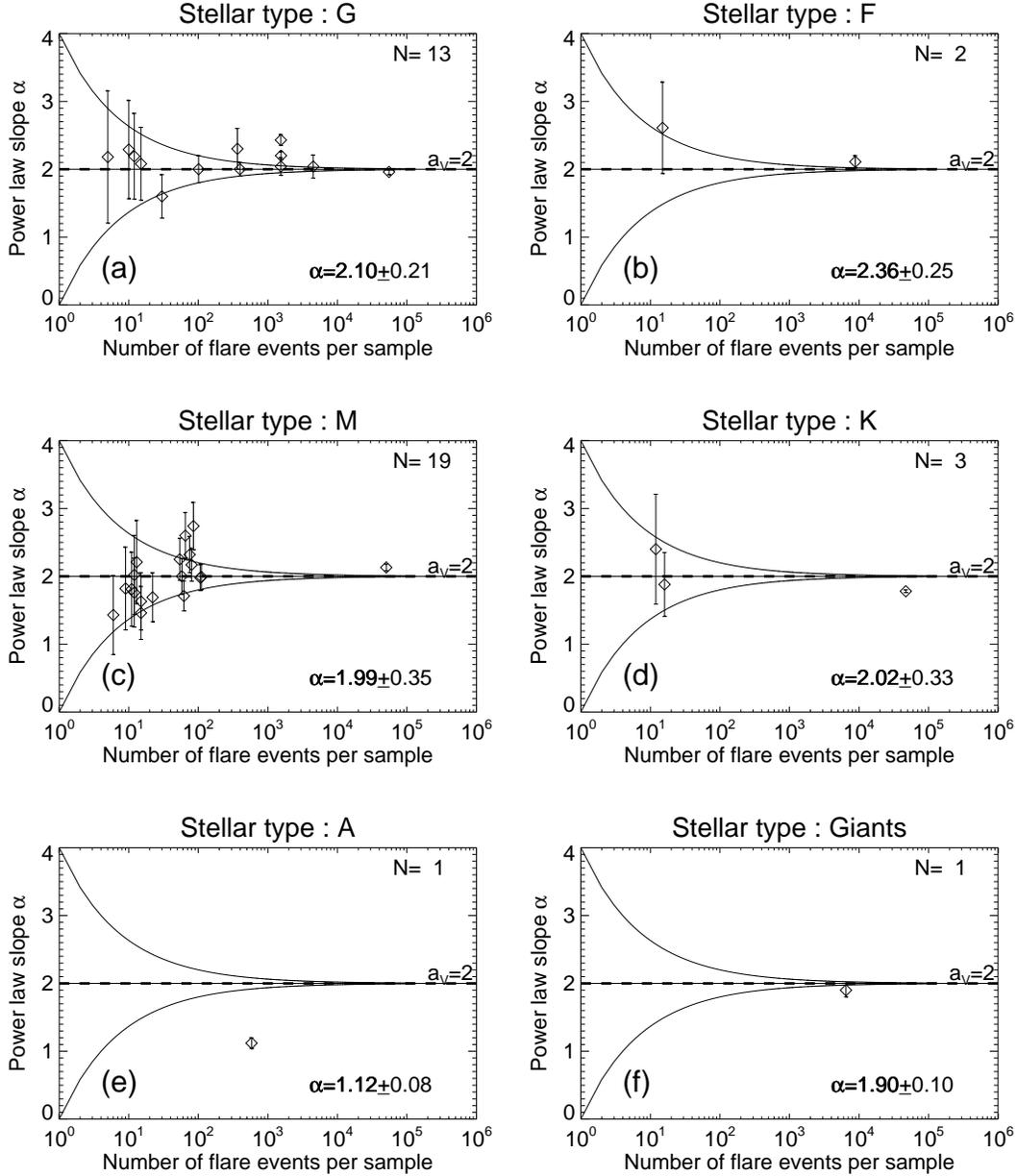}}
\caption{Means and standard deviations of power law slopes
from data sets 
gruoped by star types (A, F, G, K, M, Giants), based on the
classification of the Kepler flare catalog (Yang and Liu 2019)
and the star classification given in Tables 1 and 2.. 
The dashed horizontal lines indicate power law slopes 
predicted by a SOC model with $\alpha_V=2$, while the
curves indicate the expected uncertainties according to
Eq.~12. The number of flare data sets per stellar
spectral type varies from $N=1$ (A-type stars) to $N=13$
(G-type stars).} 
\end{figure}

\begin{figure}
\centerline{\includegraphics[width=0.8\textwidth]{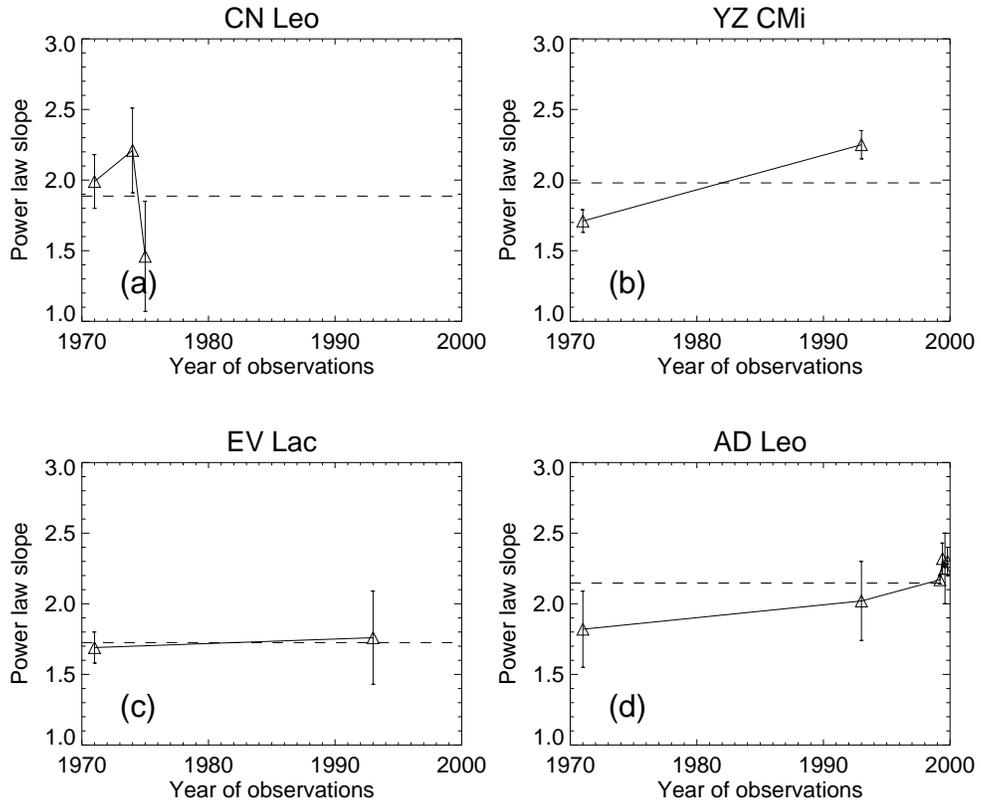}}
\caption{Time evolution of the power law slope $\alpha(t)$
of stellar flare energies observed in four stars (CN Leo, YZ CMi,
EV Lac, and AD Leo). The error bars are computed from the
standard deviation $\sigma_\alpha=\alpha/\sqrt{n}$, and the average 
value is indicated with a dashed horizontal line.}
\end{figure}

\end{document}